\documentclass{amsart}

\usepackage{amssymb}
\usepackage{graphicx}
\usepackage{epstopdf}
\usepackage{amscd}
\usepackage{graphics}
\newtheorem{Theorem}{Theorem}[section]
\newtheorem{Proposition}{Proposition}[section]
\newtheorem{Corollary}{Corollary}[section]
\newtheorem{Lemma}{Lemma}[section]

\def\proof{\par{\it Proof}. \ignorespaces}
\def\endproof{{\ \vbox{\hrule\hbox{%
     \vrule height1.3ex\hskip0.8ex\vrule}\hrule }}\par}
\newenvironment{Proof}{\proof}{\endproof}

\theoremstyle{definition}

\newtheorem{Example}[Theorem]{Example}

\theoremstyle{remark}
\newtheorem{Remark}[Theorem]{Remark}

\numberwithin{equation}{section}

\begin{document}


\title{Geometry of the Pfaff lattices}

\author{Yuji Kodama$^*$}
\address{Department of Mathematics, Ohio State University, Columbus,
OH 43210}
\email{kodama@math.ohio-state.edu}
\author{Virgil U. Pierce$^{**}$}
\thanks{$^*$Partially
supported by NSF grant DMS0404931}
\thanks{$^{**}$ Partially supported by NSF grant DMS0135308}

\address{Department of Mathematics, Ohio State University,
Columbus, OH 43210}
\email{vpierce@math.ohio-state.edu}


\begin{abstract}
The (semi-infinite) Pfaff lattice was introduced by Adler and van Moerbeke
\cite{adler:02B} to
describe the partition functions for the random matrix models of
GOE and GSE type. 
The partition functions of those matrix models are given by the Pfaffians of certain
skew-symmetric matrices called the moment matrices, and they are the $\tau$-functions of
the Pfaff lattice.
In this paper,
we study a finite version of the Pfaff lattice equation as a Hamiltonian system.
In particular,  we prove the complete integrability in the
sense of Arnold-Liouville, and using a moment map, we describe the real isospectral
varieties of the Pfaff lattice. The image of the moment map is a convex polytope
whose vertices are identified as the fixed points of the flow generated by the Pfaff lattice.
\end{abstract}

\maketitle

\thispagestyle{empty}
\pagenumbering{roman}\setcounter{page}{1}
\tableofcontents

\pagenumbering{arabic}
\setcounter{page}{1}


\section{Introduction}
This paper centers around the finite version of the Pfaff lattice hierarchy.
The Pfaff lattice is a Hamiltonian system, and has many connections to lattices of Toda type.
  Our goal is to show the complete integrability of the
 Pfaff lattice, and to describe the geometry
of the iso-level sets determined by the integrals of the lattice.

\subsection{Lie algebra splitting related to SR-factorization}
We introduce the Pfaff lattice by starting
 with a Lie algebra splitting (see \cite{adler:99, adler:02B} for the semi-infinite case),
\[
{\mathfrak{sl}}(2n,{\mathbb R})={\mathfrak k}\oplus {\mathfrak{sp}}(n,{\mathbb R})\,.
\]
Here the splitting is defined by the projections of an element $X\in{\mathfrak{sl}}(2n,{\mathbb R})$,
\begin{align*}
\pi_{\mathfrak k}(X)&=X_--J\left(X_+\right)^TJ+\frac{1}{2}(X_0-JX_0^TJ)\\
\pi_{\mathfrak{sp}}(X)&=X_++J\left(X_+\right)^TJ+\frac{1}{2}(X_0+JX_0^TJ)
\end{align*}
where $X_0$ is the $2\times 2$ block diagonal part of $X$, $X_+$ (resp. $X_-$) is
the $2\times 2$ block upper (resp. lower) triangular part of $X$, and $J$ is the $2n\times 2n$
matrix with the $2\times 2$ diagonal blocks,
\[
J_2:=\begin{pmatrix} 0 &1\\-1&0\end{pmatrix}.
\]
We denote ${\mathfrak{sp}}(n)=\{X\in{\mathfrak{sl}}(2n):\sigma(X)=X\}$ with the involution
$\sigma(X):=JX^TJ$. One finds that
$\mbox{dim}\,{\mathfrak{sp}}(n)=2n^2+n$. Each element  $Y\in{\mathfrak
  k}$ is 
expressed by a $2\times 2$ block lower triangular matrix with $2\times 2$ block diagonal
part given by $Y_0=$diag$_2(k_1I_2,\ldots,k_nI_2)$ with $\sum_{j=1}^nk_j=0$,
and $I_2$ represents the $2\times2 $ identity matrix. We also use
the $2\times 2$ block decomposition for ${\mathfrak
  g}={\mathfrak{sl}}(2n)$,
\[
{\mathfrak g}={\mathfrak g}_+\oplus {\mathfrak g}_0\oplus {\mathfrak g}_-={\mathfrak g}_{\ge0}\oplus{\mathfrak g}_-={\mathfrak g}_+\oplus{\mathfrak g}_{\le 0}\,,
\]
where ${\mathfrak g}_{\pm}$ represents the set of $2\times 2$ block upper (lower) triangular
matrices, and ${\mathfrak g}_0$ the set of $2\times 2$ block diagonal matrices.
 With the inner product on ${\mathfrak{sl}}(2n)$ defined by $\langle X,Y\rangle:={\rm tr}(XY)$,
one can define the dual spaces ${\mathfrak k}^*\cong{\mathfrak{sp}}(n)^{\perp}$ and ${\mathfrak{sp}}(n)^*\cong{\mathfrak k}^{\perp}$. With the identification ${\mathfrak g}\cong{\mathfrak g}^*$, we have
${\mathfrak{sl}}(2n)={\mathfrak k}^*\oplus {\mathfrak{sp}(n)}^*$ with the projections,
\begin{align*}
\pi_{\mathfrak k^*}(X)&=X_+-J\left( X_+\right)^TJ+\frac{1}{2}(X_0-JX_0^TJ)\\
\pi_{\mathfrak{sp}^*}(X)&=X_-+J\left(X_+\right)^TJ+\frac{1}{2}(X_0+JX_0^TJ)\,.
\end{align*}
Notice that both ${\mathfrak k}, {\mathfrak k}^{\perp}\subset {\mathfrak g}_{\le 0}\,.$
Also ${\mathfrak k}^*\cong {\mathfrak{sp}}(n)^{\perp}=\{X\in {\mathfrak{sl}}(2n):\sigma(X)=-X\}$,
which gives a vector space decomposition based on the involution
$\sigma(X)=JX^TJ$, that is,
${\mathfrak{sl}}(2n)={\mathfrak{sp}}(n)\oplus
{\mathfrak{sp}}(n)^{\perp}$. The $\mathfrak{sp}(n)^*$ is the set of
$2\times 2$ block lower triangular matrices with block diagonal part
of the form
${\rm diag}(\alpha_1,-\alpha_1,\alpha_2,-\alpha_2,\ldots,\alpha_n,-\alpha_n)$.
Note that ${\rm dim}(\mathfrak{sp}(n)^*)={\rm dim}(\mathfrak{sp}(n))=2n^2+n$.

\begin{Remark}
The splitting of Lie algebras used here is induced by the
SR-factorization of $SL(2n, \mathbb{R})$ which 
is an $Sp(n, \mathbb{R})$ version of QR-factorization.  This factorization states
that for a dense set of $A \in SL(2n, \mathbb{R})$ we can find $Q\in
{G}_\mathfrak{k}$ with
$Lie({G}_\mathfrak{k})=\mathfrak{k}$ and $P\in Sp(n, \mathbb{R})$
 such that 
$A = Q^{-1} P $ (see Theorem 3.8 in \cite{bunse:86}).  
However Theorem 3.7 in \cite{bunse:86} shows that 
the set of $A \in SL(2n, \mathbb{C})$ such that the
SR-factorization exists is not dense, for this reason we restrict our
analysis to real matrices.  Remark 3.9 in \cite{bunse:86} states that the
SR-decomposition as defined in that article is not unique; however our choice
of ${G}_\mathfrak{k}$ gives a unique decomposition.
\end{Remark} 

Let us now define a matrix $L\in {\mathfrak{sl}(2n)}^*$,
\[
L=\xi +\kappa \quad {\rm with}\quad \xi\in {\mathfrak{sp}}(n)^*\,,~~\kappa\in {\mathfrak k}^*\,.
\]
In particular, we choose $\kappa$ to be
$\kappa={\rm diag}_2(\beta_1I_2,\ldots,\beta_nI_2)+\epsilon$ with $\sum_{j=1}^n\beta_j=0$ and 
\[
\epsilon=\begin{pmatrix}
0_2 & a_1 e_-& 0_2 &\cdots &0_2\\
-a_1e_-& 0_2 & a_2e_-& \cdots &0_2\\
0_2    &  -a_2e_- & 0_2&\cdots &0_2\\
\vdots & \vdots &\vdots &\ddots &\vdots \\
0_2  &  0_2  &  0_2 &\cdots &  0_2
\end{pmatrix}.
\]
Here $I_2$ is the $2\times 2$ identity matrix, $0_2$ is the $2\times
2$ zero matrix and $e_-=\begin{pmatrix}0&0\\1&0\end{pmatrix}$. 
Then the matrix $L$ is a lower Hessenberg matrix, and the number of variables in $L$ is given by $2n^2+3n-2$.
For example, $L\in{\mathfrak{sl}}(4)$ can be written in the form,
\begin{equation}
\label{pfaffL}
L=\begin{pmatrix}
\alpha_1 & *  & 0  &  0  \\
*&-\alpha_1  &0&0\\
*&*  & \alpha_2&  *\\
*    &  *  &  *&-\alpha_2
\end{pmatrix} + 
\begin{pmatrix}
\beta_1  &  0  & 0  &  0  \\
0 &\beta_1 &a_1&0\\
0&0  &-\beta_1 &  0 \\
-a_1   &  0 &  0&-\beta_1
\end{pmatrix}\,.
\end{equation}
The number of variables is $2\cdot 2^2+3\cdot 2-2=12$. 

This choice of the matrix $L$ corresponds to the minimal nontrivial example of the Pfaff lattice
considered in this paper.

\subsection{Hamiltonian structure}

We define a Lie-Poisson bracket $\{ F,H\}_{{\mathfrak g}^*}(L)$ for functions $F,H$ on ${\mathfrak g}^*$ using the $R$-matrix $R=\frac{1}{2}(\pi_{\mathfrak k}-\pi_{\mathfrak{sp}})$ with the splitting $\mathfrak{g}=\mathfrak{k}\oplus\mathfrak{sp}(n)$:
For any $X,Y\in{\mathfrak g}$, we define the $R$-bracket
\begin{align*}
[X,Y]_{\mathfrak g}&= \displaystyle{[RX,Y]+[X,RY]}\\
   &=\displaystyle{ [\pi_{\mathfrak k}X,\pi_{\mathfrak k}Y]-[\pi_{\mathfrak{sp}}X,\pi_{\mathfrak{sp}}Y]\,.}
\end{align*}
Then the Lie-Poisson bracket is given by
\begin{equation}\label{lieP}
\{F,H\}_{{\mathfrak g}^*}(L)=\langle L, [\nabla F,\nabla H]_{\mathfrak g}\rangle\,.
\end{equation}
The Pfaff lattice for $L\in\mathfrak{g}^*$ is defined by a Hamilton equation with Hamiltonian function $H$, i.e.
\begin{align}\label{pfafflattice}
\displaystyle{\frac{dL}{dt}}&=\displaystyle{ \{H,L\}_{{\mathfrak g}^*}(L)} \\
&= \displaystyle{\pi_{\mathfrak{sp}^*}[\pi_{\mathfrak{sp}}\nabla H,L]-\pi_{\mathfrak k^*}[\pi_{\mathfrak k}\nabla H,L]\,.}
\end{align}
The solution of the Pfaff lattice can be expressed as a coadjoint orbit $Ad^*_{g(t)}$ through $L^0=\xi^0+\kappa^0$ with $\xi^0\in{\mathfrak{sp}}(n)^*, \kappa^0\in{\mathfrak k^*}$ and
$g(t)=\exp(t\nabla H)$. In general, the coadjoint orbit through $L^0$ can be written in the form,
\begin{equation}\label{coadjoint}
L(t)=Ad^*_{g(t)}(L^0)=\pi_{\mathfrak{sp}^*}\left(P\xi^0P^{-1}\right)+
\pi_{\mathfrak{k}^*}\left(Q\kappa^0Q^{-1}\right)\,,
\end{equation}
where $g(t)=Q^{-1}P$ with $P\in Sp(n,{\mathbb R})$ and $Q\in{G}_{\mathfrak k}$.
Thus the coadjoint action factors into an $Sp$-action on
  ${\mathfrak{sp}}(n)^*$ and a ${G}_{\mathfrak k}$-action on ${\mathfrak k}^*$.

If the Hamiltonian $H$ is $Sp$-invariant (i.e. $\langle L,[\nabla H,X]\rangle=0, \forall X\in {\mathfrak{sp}}(n)$), one can write $dL/dt=-[\pi_{\mathfrak k}\nabla H,L]$.
With the Hamiltonian $H_k={\frac{1}{k+1}{\rm tr}(L^{k+1})}$ (the Chevalley invariants), we define the
Pfaff lattice hierarchy,
\begin{equation}\label{lax}
\frac{\partial L}{\partial t_k}=-[B_k,L]\quad {\rm with}\quad B_k=\pi_{\mathfrak k}(L^k)\,,~ k=1,\ldots,2n-1.
\end{equation}
Then the solutions to (\ref{lax}) with initial condition $L(0)$ 
 evolve on the adjoint orbit of both $G_{\mathfrak{k}}$ and $Sp(n,\mathbb{R})$ through $L(0)$, that is,
 (\ref{coadjoint}) becomes
\[
L(t)=Ad_{Q(t)}L(0)=Ad_{P(t)}L(0)\qquad {\rm with} \quad e^{\theta(t,L(0))}=Q(t)^{-1}P(t)\,,
\]
where $\theta(t,L(0))=\sum_{k=1}^{2n-1}t_kL(0)^k$ (we will discuss more details in Section \ref{taufunctions}).

We now give a {\it canonical} Lax matrix $L$ with a form of the lower Hessenberg matrix (\ref{pfaffL}) 
in the sense that we normalize the invariant elements in $L$ under the Pfaff flow of (\ref{lax}).
Those invariants are given by the following:
\begin{Lemma}\label{invariant_lemma}
For $B\in \mathfrak{k}$ and $L$ a lower Hessenberg matrix, we have
\[
\left[ B, L\right]_{2i-1, 2i} = 0,\qquad
\sum_{j=1}^{2i-1} \left[ B, L\right]_{jj} = 0\,,\quad{\rm for}\quad i=1, 2, \dots, n.
\]
\end{Lemma}
\begin{proof}
  We first have
\begin{equation*}
\left[B, L\right]_{2i-1, 2i} = (BL)_{2i-1, 2i} - (LB)_{2i-1, 2i} 
= B_{2i-1, 2i-1} L_{2i-1, 2i} - L_{2i-1, 2i} B_{2i, 2i} \,,
\end{equation*}
where we have used the upper triangular shape of
$B$ and the lower Hessenberg form of $L$.  
One then notes that $B_{2i-1, 2i-1} = B_{2i,2i}$ from the structure of
$\mathfrak{k}$ and the result follows.  
The proof of  the other equations is a similar direct calculation
which is left to the reader.
\end{proof}

Lemma \ref{invariant_lemma} implies that $L_{2i-1,
  2i}\,,\;\mbox{for}\; i=1, 2, \dots, n,$ and 
the traces of the odd
principal parts of $L$, $\sum_{j=1}^{2i-1} L_{jj}\,,\;\mbox{for}\;
 i=1,2,\dots n$,
are invariants of the Pfaff flow (\ref{lax}).
We choose values for these expressions so that the phase space of the Pfaff lattice is
\[ Z_\mathbb{R} = \left\{ L \in \mathfrak{sl}^*(2n, \mathbb{R}) 
:\; \mbox{$L$ is lower Hessenberg,}
\; L_{2i-1, 2i} = 1, \;\mbox{and}\; \sum_{j=1}^{2i-1} L_{jj} = 0 \right\} \,.\]
The space $Z_{\mathbb R}$ with the Lie-Poisson bracket (\ref{lieP}) defines a Poisson 
manifold, and the dimension is given by
\[
{\rm dim}(Z_{\mathbb R})=4\frac{n(n-1)}{2}+n+2(n-1)=2n^2+n-2\,.
\]

The form of $L$ is a finite version of the Pfaff lattice considered in \cite{adler:02B} (see (0.12) in Theorem
0.1). 
For example in the case of $n=3$, we have
\begin{equation}\label{L}
L=\begin{pmatrix}
0 &  1  &   0 & 0 & 0 & 0 \\
*& b_1 & a_1 & 0 & 0 & 0 \\
* & * & -b_1& 1 & 0 & 0 \\
* & * & * & b_2 & a_2 & 0\\
* & * & * & * &-b_2 & 1 \\
* & * & * & * & * & 0 
\end{pmatrix}\,. 
\end{equation}
It should be noted that $L$ of this form is given by a point on a ${G}_{\mathfrak k}$-orbit through
the companion matrix $C_{\gamma}$, that is, $L=\Psi C_{\gamma}\Psi^{-1}$ for some $\Psi\in {G}_{\mathfrak k}$ and
\begin{equation}\label{companion}
C_{\gamma}=\begin{pmatrix}
0 & 1 & 0 &\cdots & 0 \\
0 & 0 & 1 & \cdots & 0 \\
\vdots & \vdots &\ddots & \ddots& \vdots\\
0 &      0       &    \cdots        &0 & 1\\
-\gamma_{2n}&\gamma_{2n-1}&\cdots & -\gamma_2& 0
\end{pmatrix}
\end{equation}
where the $\gamma_k$'s are the symmetric functions of the eigenvalues, i.e.
the characteristic polynomial $\phi_{2n}(z):={\rm det}(L-zI)=z^{2n}+\sum_{k=2}^{2n}(-1)^{k}\gamma_{k}z^{2n-k}$. We also denote the real isospectral variety of the Pfaff lattice with $\gamma_k\in\mathbb{R}$ as 
\begin{equation}\label{isospectralZ}
Z_{\mathbb R}(\gamma):=\left\{L\in Z_{\mathbb R}\,:\, F_{r,0}(L)=\gamma_r \quad {\rm for}~~
r=2,3,\ldots, 2n\right\}=\bigcap_{r=2}^{2n}\left\{F_{r,0}^{-1}(\gamma_r)\right\}\,,
\end{equation}
where $F_{r,0}(L)$ are the invariant polynomials defined by ${\rm det}(L-\lambda I)=z^{2n}+\sum_{r=2}^{2n}(-1)^rF_{r,0}(L)z^{2n-r}$
(see (\ref{Fexp})). Note that the set $\{F_{r,0}(L):r=2,\ldots,2n\}$ are equivalent to $\{H_k(L):k=1,\ldots,2n-1\}$.
The main purpose of this paper is to discuss the geometric structure of $Z_{\mathbb R}(\gamma)$
including a complete integrability of the Pfaff lattice.

 
\subsection{Outline of the paper}
The paper is organized as follow:

In Section \ref{integrability}, we discuss the integrability of the Pfaff lattice.
  While the Chevalley invariants 
\[H_k= \frac{1}{k+1} \mbox{tr}\left( L^{k+1} \right), \quad k=1,2, \dots, 2n-1, \] 
are Hamiltonians of
the lattice, they do not provide enough integrals.  We show
that the additional integrals are given by the
generating function
\[ F_L(x,y,z) = {\rm det} [ (x-y) L + y J L^T J - zI] \, . \]
Note that $F_L(1, 0, z) = \mbox{det}( L - zI)= \prod_{j=1}^{2n} (z_j
-z)$, which generates integrals equivalent to the $H_k$. 
In addition we note that $F_L$ is $Sp$-invariant, i.e. $F_{P L P^{-1}} = F_L$ for $P\in Sp(n)$.    
We follow the method of
\cite{deift:89} to construct the Casimirs and action-angle variables,
with action given by the integrals generated by $F_L(x,y,z)$.
We then define an extended Jacobian associated with the spectral curve
$\mathcal{C}(L) = \left\{ F_L(x,y,z)
=0\right\} $ which has the genus $g(\mathcal{C})=2n^2-4n+1$. The extended Jacobian is
given by the Abel-Jacobi map of a vector for which $g(\mathcal{C})$ entries are
holomorphic one-forms on $\mathcal{C}(L)$ and $2n-1$ entries are meromorphic
one-forms on $\mathcal{C}(L)$ 
(it is the presence of meromorphic one forms that makes this Jacobian
\textit{extended}).  

This integrability result 
is parallel to that of \cite{deift:89} in the sense that
the spectral curve $F_L(x,y,z) = 0$ gives the actions (and Casimirs),
while an extended Jacobian on the spectral curve gives
the angles.  A fundamental difference in the case of the Pfaff lattice is that
 the spectral curve $F_L(x,y,z) = 0$  is singular (Proposition \ref{F0}).
 The singularities provide $n$ functional relations among the coefficients in $F_L(x,y,z)$.
 We then prove that the Pfaff lattice is a complete integrable
 Hamiltonian system, with $n^2-1$ Hamiltonians and $n$ Casimirs,
 in the sense of Liouville-Arnold (Theorem \ref{integrabilitythm}).

In Section \ref{taufunctions}, we review the construction of solutions
to the Pfaff lattice by SR-factorization of $e^{tL(0)}$.
The SR-factorization decomposes $ e^{tL(0)} = Q^{-1} P$ where
$Q\in {G}_\mathfrak{k}$ and $P\in Sp(n)$.  A dense set of matrices in $SL(2n,
\mathbb{R})$ possess such a factorization \cite{bunse:86}. 
 Solutions to the Pfaff lattice are then given by $L(t) = P L(0) P^{-1}
 = Q L(0) Q^{-1}$.   
The SR-factorization leads to a Cholesky-type factorization (skew Borel decomposition) of the skew
symmetric moment matrix $M = e^{tL(0)} J e^{tL(0)^T} = Q^{-1} J Q^{-T}$.
With the embedding $L\to M$ the Pfaff flow is linear on the space of
skew symmetric matrices.  We then define the $\tau_{2k}$-function as the
Pfaffian of the $2k\times 2k$ principal part of $M$, and the set $(\tau_2,\tau_{4},\ldots,\tau_{2n-2})$  generates the
solutions of the Pfaff lattice.  We will conclude this section by working out
expressions for the moment matrix and $\tau$-functions which
correspond to discrete measure  versions of the GOE and GSE random
matrix models.  While the GOE model is a generic case of the Pfaff lattice,
the GSE model corresponds to a degenerate case of the Pfaff lattice with doubly
degenerate eigenvalues.

In Section \ref{realsolutions} we consider generic solutions of the real Pfaff lattice.
In particular, we compute the fixed points of the
Pfaff lattice flow.  These are given in the standard way by computing
the asymptotic limits as $t_1 \to \pm \infty$.  
The result is  a sorting property
similar to that found for the asymptotic limits of the Toda lattice
hierarchy.  In our case the matrix converges to a $2\times 2$ block
diagonal structure which sorts the eigenvalues by pairs from smallest
pair to largest pair as $t_1\to -\infty$ and largest to smallest for
$t_1 \to \infty$ (Theorem \ref{asymptoticL}). 
We then derive the explicit form of the fixed points of the Pfaff
lattice (Theorem \ref{fixedL}).
The total number of fixed points is shown to be $(2n)!/2^n$, which is
the order of the quotient $S_{2n}/W_P$, where $W_P$ is the parabolic
subgroup of the symmetric group $S_{2n}$ associated with $P_{2\times
  2}$.  In the case of the GSE-Pfaff lattice, the number of fixed
points is given by $n!$, the order of $S_n$.

We conclude this paper, in Section \ref{homogeneous}, by constructing the moment polytope to
characterizes the isospectral variety of the Pfaff lattice. 
We first note that the $\tau_{2k}$ can be considered as a point in
$\mathbb{P}(\wedge^{2k}\mathbb{R}^{2n})$.  We then define the moment map $\mu : {\mathbb P}(\wedge^{2k}{\mathbb R}^{2n}) \to
\mathfrak{h}_\mathbb{R}^*$ and consider the moment polytope as the
image of the moment map to describe in detail the geometry of
the isospectral variety.  We show that in the generic case, the image of the moment map
is a convex polytope described by the irreducible representation of
$SL(2n,{\mathbb R})$ on  the product $\mathbb{P}(\wedge^2{\mathbb{R}}^{2n})
\times \mathbb{P}(\wedge^4{\mathbb{R}}^{2n}) \times \dots \times
\mathbb{P}(\wedge^{2n-2}\mathbb{R}^{2n})$.  
The vertices of the polytope
correspond to the fixed points of the Pfaff lattice flow, and an open dense subset of the polytope
is given by the generic orbits of the flow (Theorem \ref{isospectralPfaff}). The boundaries of
the polytope consist of the nongeneric orbits of the Pfaff lattice
which include the solutions of the KP equation.
In the case of the GSE-Pfaff lattice, the moment polytope is given by the permutahedron
of the symmetric group $S_n$, that is, the weight polytope associated to the full flag manifold
of $SL(n,{\mathbb R})$ (Theorem \ref{GSEvariety}). This situation is similar to the case of the Toda lattice (see \cite{casian:02},
\cite{shipman:00}).


\section{Integrability}\label{integrability}

In this section, we prove that the Pfaff lattice hierarchy (\ref{lax})
is a completely integrable
Hamiltonian system. 
Deift, Li, and Tomei \cite{deift:89} 
consider the case of the generalized Toda lattice, given by the
Lie algebra splitting $ \mathfrak{sl}(n) = \mathfrak{b}_- \oplus
\mathfrak{o}(n)$, where ${\mathfrak b}_-$ is the Borel subalgebra
of lower triangular matrices with zero trace, and ${\mathfrak o}(n)$
is the set of skew symmetric matrices.  They
showed the Liouville-Arnold integrability of the Toda lattice by
finding the full family of Casimirs and action integrals, and the
conjugate angles.  While some of the integrals were found by
considering the chops (as in the case of the full Kostant-Toda lattice
\cite{ercolani:93}), Deift, Li, and Tomei also introduced the
spectral curve 
\[ F_L^{Toda}(x,y,z) = \mbox{det}\left[ (x-y) L + y L^T - zI \right]=0, \]
for the Lax matrix $L\in \mathfrak{sl}(n)^*$, which is invariant under
the $SO(n)$-adjoint action, i.e. $F_{qLq^T}^{Toda}=F_L^{Toda}$ with $q\in SO(n)$.
 The coefficients of this homogeneous polynomial give the additional
 Hamiltonians and Casimirs needed to find integrability.  In addition,
 the angles conjugate to the actions 
are a section of an extended Jacobian of this
 spectral curve.

Our proof of integrability for the Pfaff lattice equations closely follows
that in \cite{deift:89}, but with several unique structures.  
The main complications arise
from the presence of $n$ nodal singular points on the
spectral curve.

\subsection{The integrals $F_{r,k}(L)$}
Let us consider the following function,
\begin{equation}\label{curve}
F_L(x,y,z):={\rm det}[(x-y)L+yJL^TJ-zI]\,,
\end{equation}
which defines an algebraic curve on ${\mathbb{CP}}^2$ with the homogeneous coordinates
$(x:y:z)$. We denote the curve as
\[
{\mathcal C}(L) :=\{(x:y:z)\in {\mathbb{CP}}^2: F_L(x,y,z)=0\}\,.
\]
Note that the function $F_L$ is $Sp$-invariant, 
\[
F_{PLP^{-1}}=F_L\quad \forall P\in Sp(n)\,.
\]
which also implies $\langle L,[\nabla F_L,X]\rangle=0$ for any $X\in{\mathfrak{sp}}(n)$.
Since $F_L$ is invariant under the involution $\iota:(x,y,z)\to (x,x-y,-z)$, $F_L$ can be expanded in the form,
\begin{equation}\label{Fexp}
F_L(x,y,z)=\sum_{r=0}^{2n}\sum_{k=0}^{\left[r/2\right]}F_{r,k}(L)\varphi^{(r,k)}(x,y,z)\,.
\end{equation}
Here the functions $\varphi^{(r,k)}(x,y,z)$ are homogeneous polynomials of $(x,y,z)$ of degree $2n$ and 
are invariant under the involution. They are defined by
\[
\begin{cases}
\,\varphi^{(2r,k)}(x,y,z)=x^{2(r-k)}(y(x-y))^kz^{2(n-r)}, \\
\,\varphi^{(2r+1,k)}(x,y,z)=x^{2(r-k)}(2y-x)(y(x-y))^kz^{2(n-r)-1}.
\end{cases}
\]
Then because $F$ is $Sp$-invariant, the
 coefficients $F_{r,k}(L)$ provide the integrals of the Pfaff lattice.
The total number of $F_{r,k}(L)$ is $n^2+2n-1$ (note $F_{0,0}(L)=1$
and $F_{1,0}(L)=0$ are not counted).
In particular, the functions $F_{r,0}$ are equivalent to 
the Chevalley invariants which are symmetric polynomials of
the eigenvalues of $L$, i.e. $F_L(1,0,z)={\rm det}(L-zI)$. 
Other invariants $F_{r,k}$ with $k\ne 0$ provide a foliation of the isospectral variety $Z_{\mathbb R}(\gamma)$ of (\ref{isospectralZ}), which will be further discussed in Section \ref{foliation}. To determine a complete
picture of the foliation, one needs to find all the Casimirs which
define the symplectic leaves (symplectic foliation of the Poisson manifold $Z_{\mathbb R}(\gamma)$).

In order to find the independent set of the integrals, we first note that the curve (\ref{curve}) is
singular at $(0,y,z)$:
\begin{Proposition}\label{F0}
The function $F_L(0,y,z)$ is reducible, and is given by
\[
F_L(0,y,z)=\left[{\rm pf}(yJL+yL^TJ+zJ)\right]^2\,,
\]
where ${\rm pf}(M)$ indicates the Pfaffian of a skew-symmetric matrix $M$.
\end{Proposition}
\begin{Proof}
First note that the matrix in the Pfaffian,
\[
M:=yJL+yL^TJ+zJ=-J(-yL+yJL^TJ-zI)\,,
\]
is skew-symmetric. 
The determinant of a skew symmetric $2n\times 2n$ matrix is 
the square of the Pfaffian, this then proves the Proposition.
\end{Proof}
Then from Lemma \ref{F0}, we have $n$ functional relations among $F_{r,[r/2]}(L)$:
\begin{Corollary}\label{relations}
Each function $F_{n+k,[(n+k)/2]}(L)$ for $k=1,\ldots,n$ can be determined by the
first $n-1$ functions $F_{r,[r/2]}(L)$ for $r=2,\ldots,n$.
\end{Corollary}
\begin{Proof}
The function $F_L(0,1,z)$ has the expansion,
\[
F_L(0,1,z)=\displaystyle{\sum_{r=0}^{2n}  f_{r}(L)\, z^{2n-r}\,,}
\]
where $f_{r}(L)$ are defined by $ f_{r}=(-1)^{r/2}F_{r,r/2}(L)$ if $r$
 is even, and $f_{r}=2(-1)^{[r/2]}F_{r,[r/2]} (L)$ if $r$ is odd.
 From Lemma \ref{F0}, we have
\[
F_L(0,1,z)=\left(z^n+ g_2z^{n-2}+\ldots g_n\right)^2\,.
\]
where $g_k,~k=2,\ldots,n$ are calculated from the Pfaffian of the matrix $M(y,z)$.
This implies that all $g_{k}$'s are determined by $F_{r,[r/2]}$ for $r=2,\ldots,n$. Then
the next $n$ functions $F_{r,[r/2]}$ for $r=n+1,\ldots,2n$ are expressed by the quadratic
forms of $g_k$'s, that is, $F_{n+k,[(n+k)/2]}=\Phi_k(F_{2,1},\ldots,F_{n,[n/2]})$ with a polynomial
function $\Phi_k$ for $k=1,\ldots,n$.
\end{Proof}

\begin{Example}\label{exampleF} For the case of $n=2$, we have the following two relations,
\[
F_{3,1}=0,\quad\quad F_{4,2}=\frac{1}{4}F_{2,1}^2\,.
\]
For $n=3$, we have three relations:
\[
F_{4,2}=\frac{1}{4}F_{2,1}^2,\quad F_{5,2}=\frac{1}{2}F_{2,1}F_{3,1},\quad F_{6,3}=-F_{3,1}^2\,.
\]
For $n=4$, we have four relations,
\[
F_{5,2}=\frac{1}{2}F_{2,1}F_{3,1},\quad F_{6,3}=-F_{3,1}^2+F_{2,1}G_{4,2},\quad F_{7,3}=F_{3,1}G_{4,2},\quad F_{8,4}=G_{4,2}^2\,,
\]
where $G_{4,2}:=\frac{1}{2}(F_{4,2}-\frac{1}{4}F_{2,1}^2)$. 
\end{Example}
Proposition \ref{F0} indicates that the curve ${\mathcal C}(L)$ has singularities of
$n$ double points at $(0,y,z)$, and that for generic $L$ these are the only singularities,
in which case the genus of ${\mathcal C}(L)$ is given by the Clebsch formula,
\begin{equation}\label{genus}
g({\mathcal C})=\frac{1}{2}(2n-1)(2n-2)-n=2n^2-4n+1\,.
\end{equation}

Let us now find Casimirs for the coadjoint action $Ad^*_g$ for $g\in\mathfrak{g}$:
Recall that $Ad^*_g$ 
factors into an $Sp$-action and a ${G}_{\mathfrak k}$-action. We first have:
\begin{Proposition}\label{sporbit}
The coadjoint orbit generated by $Ad^*_P$ of $P\in Sp(n,{\mathbb R})$ through 
$X\in{\mathfrak{sp}(n)^*}$
is given by
\[
{\mathcal O}_{X}=\left\{ Y\in {\mathfrak{sp}}(n)^* : \hat C_k(X)=\hat C_k(Y),~k=1,\ldots,n\right\}\,,
\]
where $\hat C_k(X)$ is defined by the coefficients of $F_X(2,1,z)$,
\[
F_X(2,1,z)=z^{2n}+\sum_{k=1}^n \hat C_k(X)\, z^{2(n-k)}\,.
\]
\end{Proposition}
\begin{Proof}
With the Killing form on ${\mathfrak{sp}}(n)$, one can identify ${\mathfrak{sp}}(n)^*$ with ${\mathfrak{sp}}(n)$. An isomorphism of the identification may be given by
\[
X\in{\mathfrak{sp}}(n)^* ~\cong~ X':=X+JX^TJ\in{\mathfrak{sp}}(n)\,.
\]
Also with this identification, the coadjoint orbit ${\mathcal O}_X=\{Ad^*_P(X): P\in Sp(n,{\mathbb R})\}$ is isomorphic to the adjoint orbit ${\mathcal O}'_{X'}=\{Ad_P(X')=PX'P^{-1}:P\in Sp(n,{\mathbb R})\}$. Each point $Y$ on the orbit ${\mathcal O}_{X}$ can be then described by
\[
{\rm det}(X+JX^TJ-zI)={\rm det}(Y+JY^TJ-zI)\,,
\]
which is equivalent to $F_X(2,1,z)=F_Y(2,1,z)$. Note here that $F_X(2,1,z)=F_X(2,1,-z)$.
\end{Proof}
Thus $\{\hat C_k(L):k=1,\ldots,n\}$ forms the set of $Sp$-invariant Casimirs, and they are given by
\begin{equation}\label{casimirs}
\hat{C}_k(L)=\sum_{j=1}^{k}2^{2j}F_{2k,j}(L)\,.
\end{equation}

\begin{Remark} There is a corresponding proposition for the ${G}_{\mathfrak k}$-orbit.  
    In this case the invariants are
    generated by $F_{(k)} = \mbox{det}( L - zI)_{(k)} $ where $(A)_{(k)}$
    is the $k$'th lower-left chop of $A$;  the  matrix found by
    deleting the last $k$ rows and first $k$ columns of $A$.  The ratio of
    the first two coefficients of $F_{(k)}$ for $1\leq k \leq n-1$
    determines the
    ${G}_\mathfrak{k}$-orbit.  However, for $L$ lower
    Hessenberg, we obtain $F_{(1)} = \prod_{j=1}^{n-1} a_j$, and $F_{(k)}=
    0\,,\;\mbox{for}\; k>1$, so that the ${G}_\mathfrak{k}$-invariants are trivial.  
\end{Remark}

Thus we have $n$ relations and $n$ Casimirs in the set of $\{F_{r,k}(L)\}$ consisting of $n^2+2n-1$ functions.
The total number of Hamiltonians for the Pfaff lattice is then given by
\[
n^2-1\,.
\]
Recall that the $L$ has $2n^2+n-2$ free variables, and there are $n$ Casimirs. This means that
the Pfaff lattice (\ref{lax}) is a Hamiltonian system of degree $n^2-1$, and our goal is to
give a complete set of $n^2-1$ Hamiltonians in involution, and to construct the angle conjugate
to those Hamiltonians (action integrals).

We now discuss the commutativity of the integrals, $F_{r,k}(L)$. Let us first note:
\begin{Lemma}\label{commutation}
Let $F, G$ be functions on ${\mathfrak g}^*$.
Suppose $F$ is ${G}_{\mathfrak k}$-invariant, and $G$ is $Sp$-invariant.
Then $\{F,G\}_{\mathfrak g^*}(L)=0$.
\end{Lemma}
\begin{Proof} We obtain
\begin{align*}
\{F,G\}_{\mathfrak g^*}(L)
&=\langle L, [\pi_{\mathfrak k}\nabla F,\pi_{\mathfrak k}\nabla G]-[\pi_{\mathfrak{sp}}\nabla F,
\pi_{\mathfrak{sp}}\nabla G]\rangle\\
&= \langle L, [\nabla F,\pi_{\mathfrak k}\nabla G]-[\pi_{\mathfrak{sp}}\nabla F,\nabla G]\rangle\\
&=0.
\end{align*}
Here the first term is zero by the $G_{\mathfrak k}$-invariance of $F$, and the second
is zero by the $Sp$-invariance of $G$.
\end{Proof}
As an immediate consequence of this Lemma,
we have
\begin{Corollary}\label{chevalley}
\[
\{F_L(1,0,z),F_L(1,y',z')\}_{\mathfrak g^*}(L)=0\,.
\]
\end{Corollary}
\begin{Proof}
Note just that  $F_L(1,0,z)={\rm det}(L-zI)$ is both ${G}_{\mathfrak k}$- and
$Sp$-invariant, and $F_L(1,y', z')$ is $Sp$-invariant.
\end{Proof}

This implies $\{F_{r,0},F_{r',k'}\}_{\mathfrak g^*}(L)=0$ for $r,r'=2,3,\ldots,2n$ and $k'=0,1,\ldots,[r'/2]$.
In order to show the commutations $\{F_{r,k},F_{r',k'}\}_{\mathfrak g^*}=0$ for nonzero $k,k'$, one needs to compute
the case $\{F_L(1,y,z),F_L(1,y',z')\}_{\mathfrak g^*}$ for nonzero $y,y'$. Let us consider the function
for $(1,y,z)\notin {\mathcal C}(L)$,
\[
{H}(y,z):=\ln\,F_L(1,y,z)\,.
\]
Using $\nabla \ln {\rm det}(L)=L^{-1}$, we have, by the chain rule,
\begin{equation}\label{nabla}
\nabla {H}(y,z)=(1-y)L(y,z)^{-1}+yJL(y,z)^{-T}J\,,
\end{equation}
where we denote
\[
L(y,z):=(1-y)L+yJL^{-T}J-zI=L_y-zI
\]
 and $L^{-T}=(L^{-1})^T$.
Now we can prove
\begin{Lemma}\label{Fcommutation}
For nonzero values of $y-y'$, we have
\[
\{{H}(y,z),{H}(y', z')\}_{\mathfrak g^*}(L)=0\,.
\]
\end{Lemma}
\begin{Proof}
Since ${H}$ is $Sp$-invariant, we have
\[
\{{H}(y,z),{H}(y',z')\}_{\mathfrak g^*}(L)=\langle
L,[\nabla {H}(y,z),\nabla{H}(y',z')]\rangle.
\]
We also have the following formula with $L_y:=(1-y)L+yJL^{T}J$,
\[
L=\frac{yL_{y'}-y'L_y}{y-y'}\,.
\]
  From (\ref{nabla}) and $[L_y,L(y,z)]=0$, we have $\{{H}(y,z),{H}(y', z')\}_{\mathfrak g^*}=0$.
\end{Proof}
Corollary \ref{chevalley} and Lemma \ref{Fcommutation} now lead to
\begin{Proposition}
The integrals $F_{r,k}$ all commute, i.e.  $\{F_{r,k},F_{r',k'}\}_{\mathfrak g^*}(L)=0$ for all $r,r',k$ and $k'$.
\end{Proposition}
One can then choose the following set of integrals as the Hamiltonians for the Pfaff lattice:
\begin{equation}\label{hamiltonian}
{\mathcal H}(L):=\left\{
F_{r,k}(L): \begin{array}{lll} k=0,1,\ldots,n-2\\ r=2k+2,\ldots,2n\end{array} \right\}\,.
\end{equation}
The total number of Hamiltonians is $n^2-1$. In the next section, we construct the angles
conjugate to those Hamiltonians. Then the independence of the Hamiltonians is shown
by that of the angles.


\subsection{The angle variables conjugate to $F_{r,k}(L)$}
Before we define the angle variables conjugate to the integrals $F_{r,k}$, we
start with a preliminary Lemma, and introduce several notions on
the curve ${\mathcal C}(L)$: 
Recall the notation $L(y, z) = (x-y) L + y J L^T J - zI$, i.e. $F_L(1,y,z)={\rm det}[L(y,z)]$.
\begin{Lemma}
For generic $\mathcal{C}(L)$, we have
 ${ \rm dim}_{\mathbb C}[{\rm ker}( L(y,z) )]= 1$, 
and the kernel is given by any nonzero row of the
cofactor matrix $L(y,z)^c$.
\end{Lemma}
\begin{proof}
For $(1, y, z) \in \mathcal{C}(L)$, 
we have $\nabla F_L(1,y,z) =\nabla {\rm det}[L(y,z)]=L(y,z)^c $, from which we see 
\[ F_z(y,z):=\frac{\partial}{\partial z}F_L(1,y,z) = - \mbox{tr}\left( L(y,z)^c
  \right) \, ,\]
and 
\[ F_y(y,z):=\frac{\partial}{\partial y}F_L(1,y,z)= - \mbox{tr}\left( L(y,z)^c
 \left( J L^T J - L \right) \right) \, .\]
If $L(y,z)^c$ is the zero matrix 
then  $F_y =F_z = 0$, which, for generic $L$, only happens at $x=0$.  
If the cofactor of a matrix is
non-zero, then the dimension of the kernel of the matrix is less than
or equal to 1.  However as the determinant of the $L(y,z)$ is zero
on $\mathcal{C}(L)$, the
kernel has dimension strictly bigger than zero.  Therefore the
dimension of the kernel is 1.

One may then show that if the dimension of the kernel
of a matrix is one, then any non zero row of the cofactor matrix is a
basis for the kernel.  This is a straightforward computation using
Cramer's rule.  
\end{proof}

We define the divisor ${\mathcal D}(L)$ on ${\mathcal C}(L)$
such that with the eigenvector map $f: {\mathcal C}(L)\to {\mathbb{CP}}^{2n-1}; p\mapsto f(p)$,
with $L(y,z)f(p)=(L_y-zI)f(p)=0$,
the first element of $f(p)$ vanishes, denoted as
$f_1(p)=(e_1,f(p))=0$ for $p\in {\mathcal D}(L)$. In terms of the cofactor
$L(y,z)^c$, the ${\mathcal D}(L)$ can be expressed as
${\mathcal D}(L) \cup \iota{\mathcal{D}(L)}
= \{(L(y,z)^c)_{1,1}=0\} $, where  
$\iota{\mathcal{D}(L)}$ is the image of $\mathcal{D}(L)$ under the
involution $\iota:(x,y,z)\to(x,x-y,-z)$.
For some $y_0 \neq 0, 1$ we denote 
$P_i=(1,y_0,z_i)\in {\mathcal C}(L), \, i = 1, 2,
\dots, 2n$ (the fiber of $\mathcal{C}(L)$ over $y_0$) and
$Q_i=\iota P_i=(1,1-y_0,-z_i)$ for $i=1,\ldots, 2n$,
i.e. $L_{y_0}f(P_i)=z_i f(P_i)$ and $L_{1-y_0}f(Q_i)=-z_if(Q_i)$. Then we consider the dense set of
matrices $L$ satisfying
\begin{itemize}
\item[(a)] ${\mathcal D}(L)\cap \{F_z(y,z)=0\}=\emptyset$,
\item[(b)] $\mathcal D(L)$ is a finite set of ${\mathcal C}(L)$,
\item[(c)] ${\mathcal
  D}(L)\cap\{[(L(y,z)^c)_{1,1}]_z=0\}=\emptyset$.
\item[(d)] $\mathcal{D}(L) \cap \{ y = 0 \} = \emptyset$
\item[(e)] $\{P_i,Q_i\}\cap \mathcal D(L)=\emptyset$ for $(1,y_0,z_0)\notin{\mathcal C}(L)$,
\item[(f)] $\{P_i,Q_i\}\cap\{F_z(y,z)=0\}=\emptyset$ for
  $(1,y_0,z_0)\notin{\mathcal C}(L)$. 
\item[(g)] The only singular points of $\mathcal{C}(L)$ are at $x=0$.
\end{itemize}
Here $[(L(y,z)^c)_{1,1}]_z=\partial [(L(y,z)^c)_{1,1}]/\partial z$. The condition (c) gives the
genericity of the divisor ${\mathcal D}(L)$.
By the Riemann-Hurwitz theorem, we also note that the degree of
the divisor ${\mathcal D}(L)$, $d({\mathcal D}):={\rm deg}{\mathcal D}(L)$, is given by
\[
d({\mathcal D})=g({\mathcal C})+2n-1=2n(n-1).
\]

Now we consider a family of differentials on $\mathcal{C}(L)$ which define the angle variables conjugate to the action integral $F_{r,k}(L)$,
\begin{equation} \label{angleO} 
\omega_{r,k}(L) = g_{r,k}(y,z) \frac{dy}{F_z(y,z)} \, , \quad\mbox{for} \;
\begin{cases}
k = 0, 1, \dots, n-2, \\
r = 2k+2, \dots, 2n. \end{cases}
\end{equation}
where $g_{r,k}$ will be specified later.
By the Riemann-Roch theorem, there exist $g(\mathcal{C})=2n^2-4n+1$  holomorphic
differentials on the curve $\mathcal{C}(L)$. As will be shown below, the set of
those differentials consists of 
$n(n-2)$ odd differentials and $(n-1)^2$ even differentials with respect to the involution
$\iota:(1,y,z)\to(1,1-y,-z)$. We will use the odd holomorphic
differentials to give the angle variables conjugate to
$F_{r,k}$ for $k=1,\ldots,n-2$
and $r=2k+2,\ldots,2n$. 
For these holomorphic differentials, we use polynomial functions of $y$ and $z$ for $g_{r,k}$,
and since $\mbox{deg}(F_z) = 2n-1$, 
\[ \mbox{deg}(g_{r,k}) \leq 2n-3 . \] 
The angles generated by the odd holomorphic differentials account for
only $n(n-2)$ of the $n^2-1$ necessary angles.
The remaining $2n-1$ angles will be generated by 
meromorphic differentials $\omega_{r,0}$ for  $r=2,\ldots,2n$
 with poles at $y=0$.
This meromorphicity at $y=0$,
together with the involution on $\mathcal{C}$, explains why we
need condition (d) and $y_0 \neq 0, 1$.

Let us denote the singular locus of $\mathcal{C}(L)$ by 
\[ \mathcal{D}_\infty(L) = \left\{ (0, 1, \eta_i): 1\leq i \leq n
\right\}\,, \] and 
denote
\[ \mathcal{D}_i(L) = (1, a_i, b_i) \in \mathcal{D}(L) \, .\]
Note that $\mathcal{D}_\infty(L)$ is invariant under the coadjoint
action of $Sp(n)$, as the singular points are a feature of the curve $\mathcal{C}(L)$, while $\mathcal{D}_i(L)$ evolves.
Then we consider the integral of $\omega_{rk}$ from
$\mathcal{D}_\infty(L)$ to $\mathcal{D}(L)$:
\begin{equation*} 
\Phi_{r,k}(L)= \int_{\mathcal{D}_\infty(L)}^{\mathcal{D}(L)}
\omega_{r,k}(L) = \sum_{i=1}^{d(\mathcal{D})}
\int_{\mathcal{D}_\infty}^{\mathcal{D}_i} \omega_{r,k}(L) \, .
\end{equation*}
This defines a map, 
${\mathcal C}(L)^{d({\mathcal D})}\to {\widetilde{Jac}({\mathcal D})}$
where ${\mathcal C}^{d({\mathcal D})}$ is the $d({\mathcal D})$-symmetric product of
${\mathcal C}(L)$ and $\widetilde{Jac}(\mathcal D)$ associated with the divisor ${\mathcal D}(L)$ is an extended Jacobian defined below
(see also \cite{deift:89}). 
Then we show that $\Phi_{r,k}(L)$ 
gives the canonical angle conjugate to the integral $F_{r,k}(L)$, that is, we determine
$g_{r,k}(y,z)$ so that we have
\begin{Proposition}\label{anglePhi}
For any $F_{r',k'}\in {\mathcal H}(L)$, we have
\begin{equation}\label{PhiComm}
\left\{\Phi_{r,k},F_{r',k'}\right\}_{\mathfrak g^*}(L)=\delta_{r,r'}\delta_{k,k'}\quad
{\rm for}\quad \begin{cases}
k=1,\ldots,n-2, \\
r=2k+2,\ldots,2n\,.
\end{cases}
\end{equation}
\end{Proposition}
The total number of those $\Phi_{r,k}$ is $n^2-2n$, and the other $2n-1$ angles given in Proposition
\ref{Phi0} below are conjugate to the Chevalley
invariants $F_{r,0}$ for $r=2,3,\ldots,2n$.

To show Proposition \ref{anglePhi}, we first note:
\begin{Lemma}
Let $H(y_0,z_0)=\ln F_L(1,y_0,z_0)$ for $(1,y_0,z_0)\notin {\mathcal
  C}(L)$ and $y_0 \neq 0, 1$.
Then we have
\begin{equation}\label{dPhi1}
\left\{\Phi_{r,k},H(y_0,z_0)\right\}_{\mathfrak g^*}(L)=\frac{1}{F_L(1,y_0,z_0)}
\left(G_{r,k}(y_0,z_0)+G_{r,k}(1-y_0,-z_0)\right)\,,
\end{equation}
where $G_{r,k}(y_0,z_0)=y_0(1-y_0)(2y_0-1)g_{r,k}(y_0,z_0)$.
\end{Lemma}
\begin{Proof}
The commutator can be written as the flow generated by the Hamiltonian $H(y_0,z_0)$,
\begin{align}\label{dPhi2}
\{\Phi_{r,k},H(y_0,z_0)\}_{\mathfrak g^*}(L)&= \displaystyle{\frac{d}{dt}\Phi_{r,k}}(L)\\
   &=\displaystyle{\sum_{i=1}^{d({\mathcal D})}\frac{g_{r,k}(a_i,b_i)}{F_z(a_i,b_i)}\frac{da_i}{dt}}\,.
\end{align}
Recall that on $\mathcal{D}_i=(1,a_i,b_i = z(a_i))$, the eigenvector $f(\mathcal{D}_i)$ 
for $L_{a_i}f(\mathcal{D}_i)=b_if(\mathcal{D}_i)$ 
satisfies $(e_1,f(\mathcal{D}_i))=0$, i.e. $\mathcal{D}_i\in {\mathcal D}(L)$.
The evolution of $f(\mathcal{D}_i(t),t)$ is given by 
\begin{equation}\label{fB}
\frac{\partial}{\partial t}f(\mathcal{D}_i,t)=-B_{y}f(\mathcal{D}_i,t)\quad\mbox{and}\quad
 f(\mathcal{D}_i)=f(\mathcal{D}_i,0)\,.
\end{equation}
Here the generator $B_{y}$ can be found from the Lax equation $\frac{dL_y}{dt}=[L_y,B_y]$,
which is the isospectral condition for $L_yf=zf$:  Recall that the  Pfaff lattice with the Hamiltonian $H_0:=H(y_0,z_0)$ is $\frac{dL}{dt}=[L,\pi_{\mathfrak k}\nabla H_0]$ (see (\ref{pfafflattice})).
Then we have
\begin{align}\label{laxB}
\displaystyle{\frac{dL_y}{dt}}&= (1-y)[L,\pi_{\mathfrak k}\nabla H_0]+yJ[L,\pi_{\mathfrak k}\nabla H_0]^TJ\\
   &= [L_y,\pi_{\mathfrak k}\nabla H_0]+y[JL^TJ,J(\pi_{\mathfrak k}\nabla H_0)^TJ-\pi_{\mathfrak k}\nabla H_0]\,. \end{align}
We now compute $JL^TJ$ and $J(\pi_{\mathfrak k}\nabla H_0)^TJ-\pi_{\mathfrak k}\nabla H_0$:
Let us first note by the chain rule,
\[
\nabla H_0=(1-y_0)L(y_0,z_0)^{-1}+y_0JL(y_0,z_0)^{-T}J\,.
\]
Then we have
\[
\pi_{\mathfrak k}\nabla H_0=(1-2y_0)\pi_{\mathfrak k} \left(L(y_0,z_0)^{-1}\right)\,,
\]
which leads to
\begin{equation} \label{number1}
J(\pi_{\mathfrak k}\nabla H_0)^TJ-\pi_{\mathfrak k}\nabla H_0=(2y_0-1)\left(
L(y_0,z_0)^{-1}-JL(y_0,z_0)^{-T}J\right)\,.
\end{equation}
 We also note from $L_y=(1-y)L+yJL^TJ$ and the same with $y=y_0$ that $JL^TJ$ can be expressed as
\begin{equation} \label{number2}
JL^TJ=\frac{1-y_0}{y-y_0}L_y-\frac{1-y}{1-y_0}L_{y_0}=\frac{y_0}{y-(1-y_0)}L_y-\frac{1-y}{y-(1-y_0)}JL_{y_0}^TJ\,.
\end{equation}
Substituting (\ref{number1}) and (\ref{number2})  into  
(\ref{laxB}), we have the Lax equation 
\[
\frac{dL_y}{dt}=[L_y,B_y]\,,
\]
with
\[
B_y=\pi_{\mathfrak k}\nabla H_0+(2y_0-1)y\left(\frac{1-y_0}{y-y_0}L(y_0,z_0)^{-1}-\frac{y_0}{y-(1-y_0)}JL(y_0,z_0)^{-T}J\right)\,.
\]
Now from (\ref{fB}) and $(e_1,f(\mathcal{D}_i(t),t))=0$ for each $t$, we have $
\left(e_1,\frac{\partial f}{\partial t}+\frac{\partial f}{\partial a_i}\frac{da_i}{dt}\right)=0$ which
leads to
\[
\frac{da_i}{dt}\Big|_{t=0}=\frac{(e_1, B_{y}f(\mathcal{D}_i))}{(e_1,\partial_{y}f(\mathcal{D}_i))}\,.
\]
Since $\pi_{\mathfrak k}\nabla H_0$ is a lower triangular matrix, $(e_1,\pi_{\mathfrak k}\nabla H_0f(\mathcal{D}_i))=0$.  Then the RHS of (\ref{dPhi2}) becomes
\begin{equation} \label{number3}
\sum_{i}\frac{g_{r,k}(a_i,b_i)}{F_z(a_i,b_i)}\frac{(2y_0-1)}{(e_1,\partial_{a_i}f(\mathcal{D}_i))}
\left(e_1,\left\{\frac{a_i(1-y_0)}{a_i-y_0}L(y_0,z_0)^{-1}-\frac{a_iy_0}{a_i-(1-y_0)}JL(y_0,z_0)^{-T}J
\right\}f(\mathcal{D}_i)\right)
\,.
\end{equation}
Recall that $(e_1, f(\mathcal{D}_i)) = 0$ for $\mathcal{D}_i = (1, a_i, b_i) \in
\mathcal{D}(L)$.  Also we have that $L(y_0, z_0) f(y_0, z_j) = (z_j -
z_0) f(y_0, z_j) $ and $ J L(y_0, z_0)^T J f(1-y_0, -z_j) = L(1-y_0,
-z_0) f(1-y_0, -z_j) = - (z_j - z_0) f(1-y_0, -z_j) $ for $j=1,
\dots, 2n$. 
Then using the residue theorem, equation (\ref{number3}) can be written by
\[
\displaystyle{-y_0(1-y_0)(2y_0-1)\sum_{j=1}^{2n}\left(\frac{g_{r,k}(y_0,z_j)}{F_z(y_0,z_j)}\frac{1}{z_j-z_0}
+\frac{g_{r,k}(1-y_0,-z_j)}{F_z(1-y_0,-z_j)}\frac{1}{z_j-z_0}\right)}\,.
\]
Again by the residue theorem, we finally obtain
\[
\left\{\Phi_{r,k},H(y_0,z_0)\right\}_{\mathfrak g^*}=\frac{y_0(1-y_0)(2y_0-1)}{F_L(1,y_0,z_0)}\left(g_{r,k}(y_0,z_0)-g_{r,k}(1-y_0,-z_0)\right)\,,
\]
which is (\ref{dPhi1}).
\end{Proof}

\medskip

We now prove Proposition \ref{anglePhi}:
\begin{Proof}
With (\ref{angleO}), we choose the following polynomials for $g_{r,k}(y_0,z_0)$:

For the case $r=2s+1$, we choose,
\[
g_{2s+1,k}(y_0,z_0)=\frac{1}{2}(y_0(1-y_0))^{k-1}z_0^{2(n-s)-1}\quad{\rm for}~~ \begin{cases}
k=1,\ldots,n-2,\\ s=k+1,\ldots, n-1.\end{cases}
\]
Then we have
\begin{equation}\label{oddPhi}
\{\Phi_{2s+1,k},H(y_0,z_0)\}_{\mathfrak g^*}(L)=\frac{1}{F_L(1,y_0,z_0)}\varphi^{(2s+1,k)}(1,y_0,z_0)\,.
\end{equation}

For the case $r=2s$, we choose
\[
g_{2s,k}(y_0,z_0)=\frac{(y_0(1-y_0))^{k-1}}{2(2y_0-1)}\left(1-(4y_0(1-y_0))^{s-k}\right)z_0^{2(n-s)}\quad{\rm for}~~ \begin{cases}
k=1,\ldots,n-2,\\ s=k+1,\ldots, n, \end{cases}
\]
which gives
\begin{equation}\label{evenPhi}
\{\Phi_{2s,k},H(y_0,z_0)\}_{\mathfrak g^*}(L)=\frac{1}{F_L(1,y_0,z_0)}
\left(
\varphi^{(2s,k)}(1,y_0,z_0)-4^{r-k}\varphi^{(2s,s)}(1, y_0,z_0)\right)\,.
\end{equation}

On the other hand, using (\ref{Fexp}) for $F_L(1,y_0,z_0)$, we have
\begin{equation}\label{expPhi}
\{\Phi_{r,k},H(y_0,z_0)\}_{\mathfrak g^*}(L)=\frac{1}{F_L(1,y_0,z_0)}\sum_{r'=0}^{2n}\sum_{k'=0}^{[r'/2]}
\varphi^{(r',k')}(1, y_0,z_0)\{\Phi_{r,k},F_{r',k'}\}_{\mathfrak g^*}(L)\,.
\end{equation}
Comparing this to (\ref{oddPhi}) and (\ref{evenPhi}), we get the results of Proposition \ref{anglePhi}.
Notice that $F_{2s,s}$ is not a Hamiltonian in our list, hence the presence of the $\varphi^{(2s,s)}$
term in (\ref{evenPhi}) does not affect the statement of the Proposition.
\end{Proof}

\begin{Remark} 
The function $g_{2s, k}$ is indeed a polynomial and when inserted
into (\ref{angleO}) gives $\omega_{2r, k}$ a holomorphic one form for
$1\leq k \leq n-2$ and $k+1 \leq s \leq n$. 
\end{Remark}

In order to find the angles $\Phi_{r,0}(L)$ conjugate to $F_{r,0}(L)$, we 
consider the following integral of the meromorphic differentials $\omega_{r,0}(L)$ with a pole at $y=0$,
\[
\Phi_{r,0}(L)=\int_{{\mathcal D}_{\infty}(L)}^{{\mathcal D}(L)}\omega_{r,0}(L),\quad\quad   \omega_{r,0}(L)=\frac{m_r(y,z)+yn_r(y,z)}{y}\frac{dy}{F_z(y,z)}\,.
\]
where $m_r(y,z)$ and $n_r(y,z)$ are polynomials and $m_r(0,z)\ne 0$.
Then one can show:

\begin{Proposition}\label{Phi0}
For $F_{r',k'}\in {\mathcal H}(L)$, we have
\[
\left\{\Phi_{r,0}, F_{r',k'}\right\}_{\mathfrak g^*}(L)=\delta_{r,r'}\delta_{k',0}\quad {\rm for}~~~
r=2,3,\ldots,2n.
\]
\end{Proposition}
\begin{Proof}
As in the previous case, we compute
$\{\Phi_{r,0},H(y_0,z_0)\}_{\mathfrak g^*}(L)$ with $H(y_0,z_0)=\ln F_L(1,y_0,z_0)$ for $(1,y_0,z_0)\notin {\mathcal C}(L)$.
Then following the previous calculation, we obtain
\begin{align*}
\left\{\Phi_{r,0}, H(y_0,z_0)\right\}_{\mathfrak g^*}(L)&=\displaystyle{\frac{2y_0-1}{F_L(1,y_0,z_0)}\left[\left\{
(1-y_0)m_r(y_0,z_0)-y_0m_r(1-y_0,-z_0)\right\} \right.} \\
&\quad \left. +y_0(1-y_0)\left\{n_r(y_0,z_0)-n_r(1-y_0,-z_0)\right\}\right]\,. 
\end{align*}
Now we choose the following $m_r(y_0,z_0)$ and $n_r(y_0,z_0)$:
For $r=2s+1$, we choose
\[\begin{cases}
\,m_{2s+1}(y_0,z_0)=\displaystyle{ z_0^{2(n-s)-1} ,}\\
\,n_{2s+1}(y_0,z_0)=0,
\end{cases}\quad {\rm for}\quad s=1,\ldots, n-1.
\]
Then we have
\begin{equation}\label{oddP0}
\{\Phi_{2s+1,0},H(y_0,z_0)\}_{\mathfrak g^*}(L)=\frac{1}{F_L(1,y_0,z_0)}\varphi^{(2s+1,0)}(1,y_0,z_0)\,.
\end{equation}
For $r=2s$, we find that the polynomials
\[\begin{cases}
\,m_{2s}(y_0,z_0)= (2y_0-1)z_0^{2(n-s)},  \\
\,n_{2s}(y_0,z_0)=\displaystyle{\frac{2}{2y_0-1}\left(1-(4y_0(1-y_0))^{s-1}\right)z_0^{2(n-s)}},
\end{cases}\quad\quad {\rm for}~~~ s=1,\ldots,n,
\]
give
\begin{equation}\label{evenP0}
\{\Phi_{2s,0},H(y_0,z_0)\}_{\mathfrak
  g^*}(L)=\frac{1}{F_L(1,y_0,z_0)}\left(\varphi^{(2s,0)}(1,y_0,z_0)-4^s\varphi^{(2s,s)}(1, y_0,z_0)\right)\,.
\end{equation}
Now comparing (\ref{expPhi}) for $\{\Phi_{r,0},H(y_0,z_0)\}_{\mathfrak g^*}(L)$, we obtain the desired results as in the previous case.
\end{Proof}
Propositions \ref{anglePhi} and \ref{Phi0} provide a proof of the main theorem in this section:
\begin{Theorem}\label{integrabilitythm}
The Pfaff lattice (\ref{pfafflattice}) with $L \in Z_\mathbb{R}$ 
is a completely integrable Hamiltonian system with
$n^2-1$ Hamiltonians given in (\ref{hamiltonian}) and $n$ Casimir functions of (\ref{casimirs}).
\end{Theorem}

\subsection{Extended Jacobian}\label{extendedjack}
Here we just mention that the Pfaff flows are linearized on the extended Jacobian
defined below.
First recall that the curve ${\mathcal C}(L)$ has the genus $g({\mathcal C})=2n^2-4n+1$ (recall (\ref{genus})),
and the positive divisor ${\mathcal D}(L)$ has degree $d({\mathcal D})=2n(n-1)$.
There are $g({\mathcal C})$ number of holomorphic differentials (the Riemann-Roch theorem).
We have $n^2-2n$ differentials $\omega_{r,k}$ defined in (\ref{angleO}) for $k=1,\ldots,n-2$ and $r=2k+2,\ldots,2n$, which are {\it odd} forms with respect to the involution $\iota:(1,y,z)\to(1,1-y,-z)$.
In addition to those, we have other $(n-1)^2$ holomorphic differentials $\omega^e_{r',k'}$ 
which are {\it even} forms defined
as follows:  For $r'=2s+1$, 
\[
\omega^e_{2s+1,k}=(2y-1)(y(1-y))^{k-1}z^{2(n-s)-1}\frac{dy}{F_z(y,z)}\quad\quad
\begin{cases}  k=1,\ldots,n-2, \\ s=k+1,\ldots,n-1, \end{cases}
\]
which are just $\omega_{2s+1,k}^e=(2y-1)\omega_{2s+1,k}$, and
the total number of these is $\frac{(n-1)(n-2)}{2}$.  For $r'=2s$,
\[
\omega^e_{2s,k}=(y(1-y))^{k-1}z^{2(n-s)}\quad\quad
\begin{cases} k=1,\ldots,n-1,\\s=k+1,\ldots,n,
\end{cases}
\]
which give $\frac{n(n-1)}{2}$ differentials.
Then we have an extended Jacobian $\widetilde{Jac}({\mathcal D})$ associated with the divisor ${\mathcal D}(L)$ whose dimension is $d({\mathcal D})$,
\begin{equation}\label{jac}
\widetilde{Jac}({\mathcal D})=\left(\Phi_{r,k}(L)=\int_{{\mathcal D}_{\infty}}^{\mathcal D}\omega_{r,k}(L),~\Psi_{r',k'}(L)=
\int_{{\mathcal D}_{\infty}}^{\mathcal D}\omega^e_{r',k'}(L)\right)\,.
\end{equation}
Note here that the differentials $\omega_{r,k}(L)$ also include the
meromorphic one forms
defined for $k=0$ and $r=2,3,\ldots,2n$. 
\begin{Remark}
Since the $\Psi_{r', k'}$ in (\ref{jac}) are even with respect to the involution $\iota$, they are constant under the
flows of every Hamiltonian $F_{r, k}$.
\end{Remark}
Then the commutation relations in Propositions \ref{anglePhi} and \ref{Phi0} mean that the flows of the Pfaff hierarchy are linear on the Jacobian.

The extended Jacobian $\widetilde{Jac}({\mathcal D})$ of genus $g=g({\mathcal C})$ is well defined up to the choice of paths in the
integrals in (\ref{jac}); the path of integration may be deformed by an
element of the 
homology $H_1(\mathcal{C}\setminus \left\{ y=0 \right\}, \mathbb{Z})$.
Choose a basis of $H_1(\mathcal{C} \setminus \left\{ y=0 \right\},
\mathbb{Z})$ in the following way:  Let $\left\{ A_i, B_i
\right\}_{1\leq i \leq g}$ be a canonical basis for $H_1(\mathcal{C},
\mathbb{Z})$ with intersection product $A_i \cdot B_j = \delta_{ij}$ 
and let $\left\{ C_j \right\}_{1 \leq
  j \leq 2n}$ be (homologically trivial) 
cycles about the points $(1, 0, z_j)$ with a positive orientation,
where the $z_j$ are the eigenvalues of $L$.  
Then the Jacobian map
(\ref{jac}) is well defined up to the lattice $\Gamma$
 generated by the columns
of the $(g+2n-1)\times(2g+2n)$ period matrix 
\[ \Pi = \begin{pmatrix} P_1 & 0 \\
P_2 & N \end{pmatrix}, \]
where the rows of $\Pi$ correspond to the differential forms,
and the columns of $\Pi$ correspond to the cycles.  
$P_1$ is the $g \times 2g$ 
period matrix of the holomorphic one forms with respect to the canonical basis $\{A_i,
B_i\}$.  
$P_2$ is the $(2n-1)\times 2g$ matrix of periods of the meromorphic one
forms $\omega_{r,0}$ over the $\{A_i, B_i\}$.  While $N$ is the $(2n-1)\times 2n$
matrix where the entry 
\[  N_{2s-1, j} = - 2\pi i \frac{z_j^{2(n-s)}}{F_z(1, 0, z_j)} \]
is the residue of $\omega_{2s, 0}$ at $(1, 0, z_j)$ for
$s=1,\ldots,n$, and the entry
\[ N_{2s, j} = 2\pi i \frac{z_j^{2(n-s) -1}}{F_z(1, 0, z_j)} \]
is the residue of $\omega_{2s+1, 0}$ at $(1, 0, z_j)$ for $s=1,\ldots,n-1$.    
Note that $N = D_1 \tilde{V} D_2$, where $D_1$ and $D_2$ are diagonal
matrices, and $\tilde{V}$ is a $(2n-1)\times 2n$ Vandermonde matrix with the entries
$(\tilde V)_{i,j}=z_j^{i-1}$ for $1\le i\le 2n-1$ and $1\le j\le 2n$.
The residue theorem and a standard Vandermonde argument imply that 
$ \mbox{ker}(N) = {\rm span}_{\mathbb C}\left\{( 1, 1, \dots, 1)^T \right\} $.  
Therefore the rank of $\Gamma$ is $2g + 2n-1$.  
As a real
space, $\mathbb{C}^{g + 2n-1}/\Gamma$ has $2g + 2n-1$ compact dimensions and
$2n-1$ noncompact dimensions.

If the eigenvalues $z_j$ are all real, then the entries of $N$ are
pure imaginary.  One then checks that ${\frac{d\Phi_{r,0}}{dt_r} }=
\left\{ \Phi_{r,0}, F_{r,0} \right\} = 1$ so that $\Phi_{r,0}$
evolves only in the real direction.  Therefore as $N$ has imaginary
entries and full rank,  and $P_1$
generates a rank $2g$ lattice, the evolution of $\Phi_{r,0}$ is
entirely in a copy of $\mathbb{R}$ rather than $S^1$.   
This implies that the Chevalley flow induces a noncompact torus
action.  

\begin{figure}[t]
\includegraphics[width=10cm]{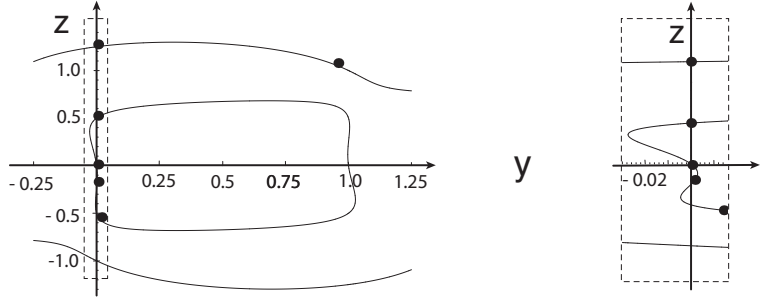}
\caption{A real section of $\mathcal{C}(L)$ with $n=3$ together with the real points of
  $\mathcal{D}(L)$ for $t\gg 0$.  In this case the six eigenvalues of $L$ are $\left\{ -1, -0.5,
  -0.25, 0, 0.5, 1.25\right\}$. The right figure is the part of the left one
 near the fiber
  $y=0$.  In particular, note that the rate at which the points in
  $\mathcal{D}(L)$ approach the eigenvalues depends on the size of the
  corresponding eigenvalue.}
\label{div:fig}
\end{figure}

Numerical calculations with $n=2$ and $3$ show that the divisor
$\mathcal{D}(L)$ 
contains some of the points $(1, 0, z_j)$, as
$t_1 \to \pm \infty$ (see Figure \ref{div:fig}).
One then notes that the meromorphic one-forms used to produce the angles
$\Phi_{r,0}$ conjugate to $F_{r,0}$ posses poles at $(1, 0, z_j)$.
The  $2n-1$  noncompact dimensions of the Jacobian produce the
fixed points of the noncompact torus actions generated by the Pfaff lattice hierarchy.

One can construct the solutions of the Pfaff lattice using the Riemann theta function through the Jacobi inversion problem.
However we will not do this here. Instead we use the matrix factorization to construct the solutions in
the next section, and will describe the (real) isospectral variety using
the moment map for the noncompact torus actions in Section \ref{homogeneous}.

\section{Matrix factorization and the $\tau$-functions}\label{taufunctions}

In this section we derive the solutions of the Pfaff lattice hierarchy
in terms of the SR-factorization of a matrix in $SL(2n)$.  We define the
$\tau$-functions whose derivatives generate solutions. In Section
\ref{homogeneous} we use the $\tau$-functions to compute the
fixed points of the Pfaff lattice by computing asymptotics as $t\to
\pm \infty$.   We conclude
this section by examining the $\tau$-functions in the 
GOE and GSE-Pfaff lattices.

As in the case of the Toda lattice, we consider the following matrix factorization for
$e^{\theta(t,L(0))}\in SL(2n,{\mathbb R})$ with the initial Lax matrix $L(0)$
and $\theta(t,L(0)):=\sum_{j=1}^{2n-1}t_jL(0)^j$,
\begin{equation}\label{qpfactorization}
e^{\theta(t,L(0))}=Q(t)^{-1}P(t)\,,
\end{equation}
where $P\in Sp(n,{\mathbb R})$ and $Q\in {G}_{\mathfrak k}$. 
The groups ${G}_{\mathfrak k}$ and $Sp(n,{\mathbb R})$ are defined by
\[
\begin{array}{llll}
{G}_{\mathfrak k}=\left\{
\begin{pmatrix}
h_1 I_2 & 0_2 & \cdots &  0_2 \\
{*}  &  h_2 I_2 & \dots &  0_2\\
\vdots &  \vdots &  \ddots &  \vdots\\
{*}  &{*} & \cdots  &  h_nI_2
\end{pmatrix} : \displaystyle{\prod_{k=1}^nh_k=1} \right\}\,,\quad {\rm dim}\,{G}_{\mathfrak k}=2n^2-n-1\,,\\
{}\\
Sp(n,{\mathbb R})=\left\{ P\in SL(2n,{\mathbb R}): PJP^T=J\right\}\,, \quad {\rm
  dim}\,Sp(n,{\mathbb R})=2n^2+n \, .
\end{array}
\]
This factorization is called the SR factorization (a symplectic version of the QR factorization),
and the set of matrices having an SR factorization is dense in $SL(2n,{\mathbb R})$,
but not in $SL(2n,{\mathbb C})$ (Theorems 3.7 and 3.8 in \cite{bunse:86}).
A standard computation starting from (\ref{qpfactorization})
 shows that the matrices $P(t)$ and $Q(t)$ satisfy
\begin{equation}\label{PQ}
\frac{\partial P}{\partial t_j}=\pi_{\mathfrak{sp}}(L^j)\, P,\quad \frac{\partial Q}{\partial t_j}=-\pi_{\mathfrak k}(L^j)\,Q\,,\quad j=1,\ldots,2n-1,
\end{equation}
and $Q(0)=P(0)=I$, the identity matrix.
The evolution of the Pfaff matrix $L(t)$ is then given by
\begin{equation}\label{laxsolution}
L(t)=Ad_{Q(t)}L(0)=Q(t)L(0)Q(t)^{-1}\,. 
\end{equation}

One should note that the factorization (\ref{qpfactorization}) is not always possible,
and in general, for some $t$, we have
\[
e^{\theta(t,L(0))}=w Q^{-1} P\quad {\rm for~some}~~ w\in S_{2n}\,,
\]
with the symmetric group $S_{2n}$. For example, consider the case $L(0)=C_0$,
the companion matrix (\ref{companion}) with $\gamma=0$, and set
$t_1=1, t_2=\frac{1}{2}, t_3=\frac{4}{3}$, which then gives
\begin{equation}\label{wSR}
 e^{\theta(t,L(0))} = \begin{pmatrix} 
1 & 1 & 1 & 2 \\
0 & 1 & 1  & 1   \\
0 & 0 &  1 & 1 \\
0 & 0 & 0  & 1 \end{pmatrix}\,. 
\end{equation}
A factorization of this matrix is given by 
\[ 
 w Q^{-1} P =\begin{pmatrix} 
0& 0 & 1 & 0 \\
0 & 0 & 0 & 1\\
1 & 0 & 0 & 0\\
0 & 1 & 0 & 0
\end{pmatrix} \begin{pmatrix}
1 & 0 & 0 & 0\\
0 & 1 & 0 & 0\\
1 & 1 & 1 & 0\\
1 & 0 & 0 & 1\\
\end{pmatrix}\begin{pmatrix}
0 & 0 & 1 & 1 \\
0 & 0 & 0 & 1 \\
1 & 1 & 0 & 0\\
0 & 1 & 0 & 0 
\end{pmatrix}
\,.\]
Here $w$ is the permutation $(1,3)(2,4)$. Note that there are other choices of
$w$, for which $w^{-1}e^{\theta(t,L(0))}$ has other SR-factorizations.
 The factorization $e^{\theta(t,L(0))}=wQ^{-1}P$ with
$w\ne id$ corresponds to a singular solution of the Pfaff lattice, and will be discussed in
a separate publication (see \cite{casian:02} for the case of the Toda lattice where the factorization is given by the Bruhat decomposition).

\subsection{Moment matrix and the $\tau$-functions}\label{moment-tau}

Let us define a $2n\times 2n$ skew-symmetric matrix
\begin{equation}\label{momentM}
M(t)=e^{\theta(t,L(0))}J e^{\theta(t,L(0))^T}=Q^{-1}JQ^{-T}\, \quad {\rm for}\quad Q\in G_{\mathfrak k}\,,
\end{equation}
which we call the moment matrix of the Pfaff lattice of (\ref{lax}).
This Cholesky-type (or skew Borel) decomposition (\ref{momentM}) represents the Gauss elimination of a skew symmetric matrix $M$.  Again one should note that in general the decomposition has the form,
for some $t$,
\[
M(t)=wQ^{-1}JQ^{-T}w^T \qquad {\rm for~some}\quad w\in S_{2n}\,.
\]
In the above example (\ref{wSR}) , the moment matrix has the form,
\[
M=\begin{pmatrix}
0 & 0 & -1 & 1 \\
  & 0 & 0 & 1\\
  &   & 0 & 1 \\
  &   &   & 0 
\end{pmatrix}\,.
\]
(In this paper, we leave blank the lower triangular part of skew-symmetric matrix.)
Notice that the element $m_{1,2}=0$, and this is the reason why one needs 
a permutation $w\neq \mbox{id}$
to have the decomposition, $M=wQ^{-1}JQ^{-T}w^T$ with $Q\in G_{\mathfrak k}$. 

Let us normalize $Q$ by a block diagonal matrix $H:={\rm diag}_2(h_1I_2,\ldots, h_{n}I_2)$ with 
$\prod_{k=1}^nh_k=1$ and $Q=H^{-1}\hat Q$ so that
\[
\hat Q=\begin{pmatrix}
I_2 & 0_2 & \cdots  & 0_2 \\
*    &    I_2 &  \cdots & 0_2\\
\vdots& \vdots &\ddots & \vdots\\
*    &    *        &\cdots & I_2
\end{pmatrix}=I_{2n}+R\,,
\]
where $R$ is a $2\times 2$ block lower triangular matrix.
This corresponds to the group factorization, ${G}_{\mathfrak k}=
{\mathcal A}\cdot \hat{G}_{\mathfrak k}$, where ${\mathcal A}$ is the set of diagonal matrices of the form of $H$ and $ \hat{G}_{\mathfrak k}$ is the set of matrices of the form of $\hat Q$. The $\hat{G}_{\mathfrak k}$ is a unipotent subgroup of $SL(2n,{\mathbb R})$, and can be identified as
the top cell (open dense subset) of the flag variety $SL(2n,{\mathbb R})/P_{2\times 2}$ where $P_{2\times 2}$ is
the parabolic subgroup of $2\times 2$ block upper triangular matrices.
This implies that the canonical morphism $\psi :SL(2n,{\mathbb R})\to SL(2n,{\mathbb R})/P_{2\times 2}$ gives an open immersion for the restriction of $\psi$ on $\hat{G}_{\mathfrak k}$, i.e.
\begin{equation}\label{immersion}
\psi\,: \, \hat{G}_{\mathfrak k}\, \overset{\cong}\longrightarrow \, {\Sigma}_{id}\,,
\end{equation}
where ${\Sigma}_{id}$ is the top cell of the flag $SL(2n,{\mathbb R})/P_{2\times 2}$.

With the definition of $\hat Q$, i.e. $\hat Q=HQ$, we have
\begin{Lemma}\label{LemmaQ}
The matrix $\hat Q$ satisfies
\[
\frac{\partial \hat Q}{\partial t_j}=-\hat B_j\hat Q\,,
\]
where $\hat B_j$ is a $2\times 2$ block lower triangular matrix given by
\[
\hat B_j=H\left((L^j)_--J(L^j)_+^TJ\right)H^{-1}\,.
\]
Also the matrix $H$ satisfies
\[
\frac{\partial H}{\partial t_j}=\frac{1}{2}\left((L^j)_0-J(L^j)^T_0J\right)H\,.
\]
\end{Lemma}
\begin{Proof}
 From (\ref{PQ}), we have
 \[
 \hat Q^{-1}\frac{\partial \hat Q}{\partial t_j}=-H \pi_{\mathfrak k}(L^j)H^{-1}+\frac{\partial H}{\partial t_j}H^{-1}\,.
 \]
 Since $\hat Q^{-1}\frac{\partial \hat Q}{\partial t_j}\in {\mathfrak g}_-$ and $\frac{\partial H}{\partial t_j}H^{-1}\in{\mathfrak g}_0$, calculating the projection $\pi_{\mathfrak{k}}(L^j)$ the statement follows.
 \end{Proof}

 The elements $h_k$ in $H$ can be calculated from the Pfaffians of the matrix $M$:
Let $M_{2k}$ be the $2k\times 2k$ principal part of $M$, the $2k\times 2k$ upper-left block of $M$.
We then define the $\tau$-function $\tau_{2k}$ as the Pfaffian of $M_{2k}$, 
\begin{equation}\label{tau}
\tau_{2k}={\rm pf}(M_{2k})={\rm
  pf}((H^2J)_{2k})=\prod_{j=1}^{k}h_j^2\,,
\quad k=1,\ldots,n.
\end{equation}
where $(H^2J)_{2k}={\rm diag}_2(h_1^2J_2,\ldots,h_k^2J_2)$.
The explicit form of the Pfaffian of a matrix 
in (\ref{tau}) is given in (\ref{pfafftau}).

We wish to describe the fixed points of the Pfaff lattice by computing
the limits of the flow as $t_j \to \pm \infty$; however for the Pfaff
variable $L\in Z_\mathbb{R}(\gamma)$ the flows are unbounded in those
limits  (the Chevalley flows correspond to non-compact torus
actions on the extended Jacobian, see
Section \ref{extendedjack}).  We then normalize these flows
by conjugating $L$ by $H$: define $\hat L$ as
\[
\hat L=HLH^{-1}=\hat Q L(0)\hat Q^{-1}\, ,
\]
 from which the elements $a_k$ in the super diagonal of $L$ are given by 
\begin{equation}\label{a}
a_k=a_k(0)\frac{h_{k+1}}{h_{k}}=a_k(0)\frac{\sqrt{\tau_{2k+2}\tau_{2k-2}}}{\tau_{2k}}\,,\quad k=1,\ldots,n-1\,.
\end{equation}
All other entries of $L$ can also be described by the set of $\tau_{2k}$
functions as will be 
shown in Section \ref{realsolutions} (see also \cite{adler:02}).  In this paper, we consider the case where $a_k(0)$ are nonzero,
and we normalize $\hat{L}$ by a conjugation so that the elements on the super diagonal are all 1.
A Pfaff lattice having $a_k(0)=0$ for some $k$ splits into smaller lattices (sub-systems) which 
appear as the boundaries of the isospectral variety of the generic flow. We then define the isospectral variety
of the normalized Pfaff lattice with matrix variable $\hat{L}$ as
\begin{equation}\label{hatphase}
\hat{Z}_{\mathbb R}(\gamma)=\left\{\,
\hat{L}\in Z_{\mathbb R}(\gamma)\,:~\hat{L}_{k,k+1}=1,~~k=1,\ldots,n-1\right\}\,.
\end{equation}
Now the asymptotic matrices of $\hat{L}(t)$ as $t_j\to\pm\infty$ are
bounded, lie in $\hat Z_{\mathbb{R}}(\gamma)$,  and
 are fixed points of the Pfaff lattice.

\begin{Remark}\label{todaM}
In the case of tridiagonal symmetric Toda lattice, the symmetric moment matrix and $\tau$-functions are
defined as follows: Let $L_T(t)$ be an $n\times n$ Lax matrix for the Toda lattice
of $\mathfrak{sl}(n,\mathbb{R})$. We have the QR
factorization,
\[
e^{\theta(t,L_T(0))}=r(t)^{-1}q(t)\quad r(t)\in B_+,~q(t)\in O(n)\,,
\]
where $B_+$ is the Borel subgroup of upper triangular matrices.
The (symmetric) moment matrix $M_T$ is then defined by
the Cholesky factorization,
\[
M_T=e^{2\theta(t,L_T(0))}=(r^{-1}q)(q^Tr^{-T})=r^{-1}r^{-T}\,.
\]
The $\tau$-functions are defined by
\begin{equation}\label{todatau}
\tau_k^{Toda}={\rm det}((M_T)_k)\,,
\end{equation}
where $(M_T)_k$ is the $k\times k$ upper left submatrix of the $M_T$ matrix.
We also obtain similar formulae to (\ref{a}) for the off diagonal elements in $L_T$ matrix,
which are also denoted as $a_k$ (see e.g. \cite{casian:04}).
\end{Remark}

We note that $\tau_{2k}$ can be parametrized by 
a constant skew-symmetric matrix determined by
the initial matrix $L(0)$, which we call the $B$-matrix, as follows:
With the normalization $a_k(0)=1$ (see above), the elements in the super-diagonal of $L(0)$ 
are all 1. One can then write an eigenmatrix $\Phi_0$ for $L(0)$, in the form
$\Phi_0=\Psi_0V$ where $\Psi_0\in\hat{G}_{\mathfrak k}$ with
diag$(\Psi_0)=I$, and the Vandermonde matrix $V$ is given by 
\[
V=\begin{pmatrix}
1   &  1  &   \cdots  &  1 \\
z_1 & z_2 & \cdots & z_{2n}\\
\vdots &\vdots & \ddots & \vdots\\
z_1^{2n-1}&z_2^{2n-1}&\cdots & z_{2n}^{2n-1}
\end{pmatrix} .
\]
Then we have 
\begin{align*}
M&=\displaystyle{\Phi_0e^{\theta(t,\Lambda)}\Phi_0^{-1}J\Phi_0^{-T}e^{\theta(t,\Lambda)}\Phi_0^T}\\
  &= \displaystyle{\Psi_0 E(t,\Lambda)B E(t,\Lambda)^T\Psi_0^T}\,.
  \end{align*}
Here the $B$-matrix is defined by
\begin{equation} \label{Bform}B=\Phi_0^{-1}J\Phi_0^{-T}\,,\end{equation}
 $\Lambda=$diag$(z_1,\ldots,z_{2n})$ (i.e. $L(0)\Phi_0=\Phi_0\Lambda$ with the eigenvalues $z_i$), and $E(t,\Lambda)=Ve^{\theta(t,\Lambda)}=e^{\theta(t,C_{\gamma})}V$ is given by
the Wronskian matrix of $\{E_1(t),\ldots, E_{2n}(t)\}$ with respect to the $t_1$ variable, i.e.
\begin{equation}\label{wronskianE}
E(t,\Lambda)=\begin{pmatrix}
E_1 & E_2 &\cdots & E_{2n} \\
E_1'&E_2' & \cdots & E_{2n}'\\
\vdots &\vdots &\ddots &\vdots \\
E_1^{(2n-1)}&E_2^{(2n-1)}&\cdots &E_{2n}^{(2n-1)}
\end{pmatrix}\,,
\end{equation}
with $E_k(t)=\exp\theta(t,z_k)$ for $k=1,\ldots,2n$, and $E_k^{(j)}=\frac{\partial^j E_k}{\partial t_1^j}=z_k^jE_k$.
Note also that the matrix $E(t,\Lambda)$ satisfies the linear equations,
\begin{equation}\label{Eequation}
\frac{\partial}{\partial t_k}E(t,\Lambda)=C_{\gamma}^kE(t,\Lambda)\,,\quad\quad E(0,\Lambda)=V\,,
\end{equation}
where we have used $V\Lambda=C_{\gamma}V$ with the companion matrix $C_{\gamma}$.
Since $\Psi_0$ does not affect the value of $\tau$-functions, it is convenient to define 
$\tilde M:=\Psi_0^{-1}M\Psi_0^{-T}=EBE^T$. The $\tilde M$ then satisfies
\begin{equation}\label{Mequation}
\frac{\partial \tilde M}{\partial t_k}=C_{\gamma}^k\tilde M+\tilde M(C_{\gamma}^k)^T\,,\quad \tilde M(0)=VBV^T\,.
\end{equation}
This equation can be considered as a linearization of the Pfaff lattice on the
space of skew symmetric invertible matrices as described in Section \ref{homogeneous}. The $B$-matrix then gives the initial
point of the flow, hence it determines the symplectic leaves of the foliation with the Hamiltonians,
$F_{r,k}(L)$. 

We remark here that the Pfaff flow can be linearized on $SL(2n,{\mathbb R})/Sp(n,{\mathbb R})$ as an image of
the companion embedding, $c_{\gamma}: Z_\mathbb{R}(\gamma) \to SL(2n,{\mathbb R})/Sp(n,{\mathbb R})$, that is, we have
the commuting diagram,
\[
\begin{CD}
L(0) @ > c_{\gamma}> >  \Psi_0^{-1}\mod Sp(n,{\mathbb R})\\
@ V  VV @ VV V \\
L(t)@ > c_{\gamma}> > e^{\theta(t,C_{\gamma})}\Psi_0^{-1} \mod Sp(n,{\mathbb R})
\end{CD}
\]
Here we have used $L(0)=\Psi_0C_{\gamma}\Psi_0^{-1}$ and $L(t)=Ad_{Q(t)}L(0)$ with
$e^{\theta(t,L(0))}=Q(t)^{-1}P(t)$.
 Note here that the Pfaff flow on $SL(2n,{\mathbb R})/Sp(n,{\mathbb R})$ satisfies
the same equation as $E(t,\Lambda)$ in (\ref{Eequation}).

Now we can express the $\tau$-functions in the form with $2k\times 2k$ upper-left block of $\tilde M$,
\begin{equation}\label{pfafftau}
\tau_{2k}={\rm pf}({\tilde M_{2k}})=\sum_{I_{2k}}~
\sigma(i_1,j_1,\ldots,i_{k},j_k)\,m_{i_1,j_1}m_{i_2,j_2}\cdots
m_{i_k,j_k} \, .
\end{equation}
where the sum is taken over the set $I_{2k}=\{1=i_1<\ldots<i_k\le 2k,~i_s<j_s,~s=1,\ldots,k\}$.
The coefficient $\sigma(i_1,j_1,\ldots,i_k,j_k)$ is a parity of the permutation, i.e.
\[
\sigma:={\rm sign}\begin{pmatrix}
1 & 2 & \cdots & 2k-1&2k\\
i_1&j_1&\cdots&i_k&j_k
\end{pmatrix} \, .
\]
For example, we have
\begin{align*}
\tau_2&=m_{1,2}, \\
\tau_4&=m_{1,2}m_{3,4}-m_{1,3}m_{2,4}+m_{1,4}m_{2,3}.
\end{align*}
The $\tau$-functions are also obtained as the exterior products of the bi-vector
$\displaystyle{\Omega_2:=\sum_{1\le i<j\le 2n}m_{i,j} e_i\wedge e_j}$, i.e.
\[
\tau_{2k}=\langle \wedge^{k}\Omega_2, e_1\wedge\cdots\wedge e_{2k}\rangle\,,
\]
where $\langle\cdot,\cdot\rangle$ is the usual inner product on $\wedge^{2k}{\mathbb R}^{2n}$.
This indicates that the nonzero $\tau$-functions give the generic orbits of the Pfaff lattice,
that is, those orbits are on the top cell of the homogeneous space  $SL(2n)/Sp(n)$
(see Section \ref{homogeneous}).

The entry $m_{i,j}$ of the moment matrix $\tilde M=EBE^T$ is explicitly expressed as
\begin{equation}\label{Mentries}
\begin{array}{lll}
m_{i,j}&=&\displaystyle{\sum_{1\le k<l\le 2n}} b_{k,l}\left|\begin{matrix}
E_k^{(i-1)} & E_k^{(j-1)}\\
E_l^{(i-1)} &E_l^{(j-1)}
\end{matrix}\right|\\
&{}&\\
&=&\displaystyle{\sum_{1\le k<l<\le 2n}}b_{k,l}(z_kz_l)^{i-1}(z_l^{j-i}-z_k^{j-i})E_kE_l\,.
\end{array}
\end{equation}
      From this expression, we have
\begin{equation}\label{momentequation}
\frac{\partial m_{i,j}}{\partial t_k}=m_{i+k,j}+m_{i,j+k}\quad k=1,2,\ldots
\end{equation}
This, of course,  is the same as (\ref{Mequation}), but notice that (\ref{Mequation}) includes
the characteristic polynomial $\phi_{2n}(z)=z^{2n}+\sum_{k=2}^{2n}(-1)^k\gamma_kz^{2n-k}=0$.

Then it can be shown that the $\tau$-functions satisfy the Hirota bilinear form \cite{hirota:91},
\begin{equation}\label{dkptau}
(-4D_1D_3+D_1^4+3D_2^2)\,\tau_{2k}\cdot\tau_{2k}=24\tau_{2k-2}\tau_{2k+2}\,,
\qquad k=1,2,\ldots,n-1,
\end{equation}
with $\tau_0=1$. Here $D_k$ is the Hirota derivative with respect to $t_k$, $D_kf\cdot g:=(\partial_{t_k}-\partial_{t_k'})f(t)g(t')|_{t=t'}$. 
Note here that the first equation in (\ref{dkptau}) with $\tau_{2k+2}\tau_{2k-1}=0$ is 
the usual KP equation, and the $\tau_{2k}$ for the KP equation is given by
the Wronskian form (for example, see \cite{kodama:04}).
One should note in (\ref{dkptau}) that $\tau_{2k}$ are all generated by $\tau_2$.
This is similar to the case of Toda lattice where the $\tau$-functions $\tau^{Toda}_k$
satisfy 
\[
D_1^2\tau^{Toda}_k\cdot\tau^{Toda}_k=\tau^{Toda}_{k-1}\tau^{Toda}_{k+1}\,,\qquad k=1,\ldots,n-1,
\]
with $\tau_0^{Toda}=1$. From this recursive equation,  one can show that $\tau^{Toda}_k$ are given by the Wronskian (Hankel) determinants,
\[
\tau^{Toda}_k=Wr(\tau^{Toda}_1,(\tau^{Toda}_1)',\ldots,(\tau^{Toda}_1)^{(k-1)})\,,
\]
where $(\tau^{Toda}_1)^{(j)}=\partial^j\tau^{Toda}_1/\partial t_1^j$. This expression of the $\tau$-function then agrees with (\ref{todatau}) in Remark \ref{todaM}.

\begin{Remark}
The system (\ref{dkptau}) has been proposed as a first member of the DKP hierarchy in \cite{jimbo:83}.
In \cite{hirota:91}, Hirota and Ohta introduced the system as an extension of the KP equation
and called it the coupled KP equation. The system (\ref{dkptau}) was rediscovered as the Pfaff lattice describing the partition function of a skew-symmetric matrix models in \cite{adler:99, adler:02B, kakei:99}.
In particular, Adler et al in \cite{adler:02B} discussed the lattice structure of the system 
and formulated it as the Pfaff lattice with the connection to the Toda lattice. The system (\ref{dkptau})
was also found as a charged BKP hierarchy describing an orbit of some infinite-dimensional Clifford group action in \cite{kac:98}.
\end{Remark}


\subsection{Foliation of the phase space by $F_{r,k}(L)$}\label{foliation}

The integrals $F_{r,0}(L)$ (the Chevalley invariants) define the
 isospectral manifolds $Z_\mathbb{R}(\gamma)$.  
We have shown that the $Z_\mathbb{R}(\gamma)$ embeds into
the space of skew-symmetric invertible matrices.  This space is then
foliated by the additional integrals $F_{r,k}(L)$  for $k>0$ (found in Section
\ref{integrability}).  
Our point of view is that the $B$-matrix of (\ref{Bform}) gives coordinates on the space of
skew-symmetric matrices.  
We can then write the constants $F_{r,k}(L)$ in terms of the eigenvalues
$\{z_i:i=1,\ldots,2n\}$ and 
the $B$-matrix: We have 
\begin{Proposition}\label{levelset}
\[
F_{L}(x,y,z)=\frac{1}{{\rm det}\,(B)}\,{\rm det}[(x-y)\Lambda B-yB\Lambda-zB].
\]
\end{Proposition}
\begin{Proof} Using $L=QL(0)Q^{-1}$ and $M=Q^{-1}JQ^{-T}$, we have
\begin{align*}
F_L(x,y,z)&={\rm det}[(x-y)L+yJL^TJ-zI] \\
&= {\rm det}[(x-y) QL(0)Q^{-1}+yJQ^{-T}L(0)^TQ^TJ-zI]\\
&= {\rm det}[(x-y)L(0)-yML(0)^TM^{-1}-zI]\,.
\end{align*}
By factoring $M$ out and setting $t=0$, this can also be expressed as
\begin{align*}
 {\rm det}(M(0))\,F_L(x,y,z)&={\rm det}[(x-y)L(0)M(0)-yM(0)L(0)^T-zM(0)]\\
&={\rm det}[(x-y)\Lambda B-yB\Lambda -zB]\, {\rm det}(\Phi_0)^2\,.
\end{align*}
In the last step, we have used $L(0)\Phi_0=\Phi_0\Lambda$, $M(0)=\Phi_0B\Phi_0^{-T}$.
Then noting ${\rm det}(M(0))={\rm det}(B)\,{\rm det}(\Phi_0)^2$, we obtain the result.
\end{Proof}
We denote $\mbox{pf}(i_1, i_2, \dots, i_k)$
as the Pfaffian of the $k\times k$ skew symmetric matrix found by
taking the $i_1, i_2, \dots, i_k$ rows and columns of the $B$-matrix, e.g.
${\rm pf}(B)={\rm pf}(1,2,\ldots,2n)$.  We note that ${\rm pf}(B)=[{\rm det}(V)]^{-1}=[\prod_{i<j}(z_j-z_i)]^{-1}$
 since $1={\rm det}( M(0))={\rm det}(\Phi_0)^2{\rm det}(B)$. 

\begin{Example} The case of $n=2$: The $F_{r,k}(L)$ are expressed as
\begin{align*}
F_{2,0} &= z_1z_2+z_1z_3+z_1z_4+z_2z_3+z_2z_4+z_3z_4,   \\
F_{2,1}
&= 2 \left( b_{1,2} b_{3,4} (z_1+z_2)^2 - b_{1,3} b_{2,4} (z_1 +
z_3)^2 + b_{1,4} b_{2,3} (z_1+z_4)^2 \right)/{\rm pf}(B),   \\
F_{3,0}&=(z_1+z_2)(z_1+z_3)(z_1+z_4), \\
F_{3,1}&=0, \\
 F_{4,0}&=z_1z_2z_3z_4,  \\
F_{4,1} &=-\left(
b_{1,2}b_{3,4}(z_1+z_2)^2(z_1z_2+z_3z_4)-b_{1,3}b_{2,4}(z_1+z_3)^2(z_1z_3+z_2z_4)\right. 
\\
&\phantom{=} \left.+b_{1,4}b_{2,3}(z_1+z_4)^2(z_1z_4+z_2z_3)\right)/{\rm pf}(B),  
\end{align*}
with $F_{4,2}=\frac{1}{4}(F_{2,1})^2$ (see Example \ref{exampleF}). Here $b_{i,j}$ are the elements of the $B$-matrix. Recall that $F_{r,0}$ for $r=2,3,4$ are
the Chevalley invariants which do not depend on the $B$-matrix,  and $\hat C_1=4F_{2,0}+F_{2,1}$, $\hat C_2=16F_{4,0}+4F_{4,1}+F_{4,2}$
are the Casimirs. This
gives $F_{2,1}=\hat{c}_1\,, \; F_{4,2} = \hat{c}_2\,,$ and in addition we
have $\mbox{pf}(B)=b_{1,2}b_{3,4}-b_{1,3}b_{2,4}+b_{1,4}b_{2,3}=\hat{c}_3$, where $\hat{c}_1$, $\hat{c}_2$, and $\hat{c}_3$
are constants.  This system of equations is linear in $b_{1,2}b_{3,4}$,
$b_{1,3}b_{2,4}$, and $b_{1,4}b_{2,3}$. For generic $z_1$, $z_2$,
$z_3$, and $z_4$ it is a non-singular
system.  
The foliation of the isospectral variety $Z_{\mathbb R}(\gamma)$ is therefore
 parametrized by
\[
c_1=b_{1,2}b_{3,4},\quad c_2=b_{1,3}b_{2,4},\quad c_3=b_{1,4}b_{2,3}\,,
\]
where $c_1$, $c_2$, and $c_3$ are constants.
Note that as the $\mbox{pf}(B)\neq 0$ not all $c_i$'s are zero.  
This foliation is related to the cell decomposition of the space of
non-singular skew-symmetric $2n\times 2n$ matrices.   

\vskip 0.3cm

 The case of $n=3$:
We find for instance that
\begin{equation}\label{F21_n3}
F_{2,1} = \sum_{1\leq i < j \leq 2n} 
(-1)^{i+j+1} b_{i,j} \mbox{pf}(1, 2, \dots, \hat{i}, \dots, \hat{j}, \dots,
2n)  (z_i+z_j)^2/ {\rm pf}(B),
\end{equation}
where $\hat{i}$ means that these elements are skipped in the list. 
Our hypothesis is that this formula is true for general $n$.  
\end{Example}

\begin{Remark}
  One
should compare these results with the formula computed in
\cite{ercolani:93} ((23) in p.197) for the $k$-chop integrals of the full Kostant Toda
lattice which are found to be rational
functions of the Pl\"ucker coordinates $\pi_J$ and $\pi_J^*$
with $J\subset\{1,\ldots,n\}$ and $|J|=k$ on the flag manifold $SL(n,{\mathbb C}) / B_+$
with the Borel subgroup $B_+$.  
\end{Remark}


\subsection{Examples from the matrix models}
The Pfaff lattice was introduced as an integrable lattice whose $\tau$-functions are
the partition functions for the matrix models of  the Gaussian orthogonal ensemble (GOE)
and Gaussian symplectic ensemble (GSE) \cite{adler:99}.
The partition function of a random 
matrix model is given by the matrix integral,
\begin{equation} \label{taugen}
\tau(t_1,t_2,\ldots)=\int_{\mathcal M_N}e^{{\rm tr}(-V(X)+\theta(t,X))}\,
dX\,, 
\end{equation}
with Haar measure $dX$ on the matrix ensemble ${\mathcal M}_N$, and
some potential function $V(X)$, e.g. $V(X)=\frac{1}{2}{\rm tr} (X^2)$.
In the case of GOE, the ${\mathcal M}_N$ is given by the set of $N\times N$ real symmetric matrices,
and in the case of GSE, ${\mathcal M}_N$ is the set of $N\times N$ self-dual Hermitian matrices with
quaternionic entries.
The moment matrices for both cases are given by  skew-symmetric
matrices  and the $\tau$-functions (\ref{taugen}) are Pfaffians of the
moment matrices \cite{adler:99}.
In the following two subsections, we give explicit forms of the moment matrix and the $\tau$-functions for
those models.
\subsubsection{The matrix model of GOE}
The moment matrix associated to GOE is given by
\[
m_{i,j}=\int\int_{{\mathbb R}^2} x^i y^j\epsilon(x-y)\rho_t(x)\rho_t(y)\,dxdy\,,
\]
where $\epsilon(x)={\rm sign}(x)$, and $\rho_t(x)=\exp(-V(x)+\theta(t,x))$.

For our finite dimensional case, fix the eigenvalues of the $L$ of the Pfaff
lattice, $z_1 < z_2 < \dots < z_{2n}$,  set $dz$ to be $\sum_j^{2n}\delta(z-z_j)dz$. Then we have:
\begin{equation}\label{mGOE}
m_{i,j}=\sum_{1\le k<l<\le 2n}(z_kz_l)^{i-1}(z_l^{j-i}-z_k^{j-i})E_kE_l\,, \;\mbox{for}\; 1\le i<j\le 2n\,,
\end{equation}
where $E_k = \exp( - V(z_k) + \theta(t, z_k))$.
This is obtained from (\ref{Mentries}) with the choice of $b_{k,l}={\rm sign}(l-k)$ for the $B$-matrix
(\ref{Bform}). The $\tau$-functions
are then given by
\begin{equation}\label{tauGOE}
\tau_{2k}={\rm pf}(M_{2k})=\sum_{1\le i_1<\cdots<i_{2k}\le 2n}\left(\prod_{j<l}|z_{i_l}-z_{i_j}|\right)E_{i_1}\cdots E_{i_{2k}}\,.
\end{equation}
One should note that the coefficient of each exponential term in
(\ref{tauGOE}) is
positive, and this gives a generic orbit of the Pfaff lattice on the
homogeneous space $SL(2n)/Sp(n)$ as will be discussed in Section
\ref{homogeneous}. 

\subsubsection{The matrix model of GSE \label{GSEmatrixmodel}}
The moment matrix associated to GSE is given by
\[
m_{i,j}=\int_{\mathbb R}\{z^i,z^j\}\rho_t(z)^2dz=(j-i)\int_{\mathbb R}z^{i+j-3}\rho_t(z)^2\,dz\,.
\]
with $\{f,g\}=fg'-gf'$.
For the finite dimensional case, we have
\begin{equation}\label{mGSE}
m_{i,j}=(j-i)\sum_{k=1}^{n}z_k^{i+j-3}E_k^2\,,\;\mbox{for}\; 1\le i<j\le 2n\,.
\end{equation}
To obtain this from (\ref{Mentries}), we first take $b_{2k-1,2k}=\frac{1}{z_{2k}-z_{2k-1}}$ and
$b_{i,j}=0$ for all other $(i,j)$. Then  we take the limit $z_{2k}\to z_{2k-1}$ giving
doubly degenerate eigenvalues. In the formula, we have relabeled the indices of $z_k$ (i.e. $k$ for $2k-1$). Note that the matrix $L$ in this case has $n$ double eigenvalues.
The $\tau$-functions are then given by
\begin{equation}\label{tauGSE}
\tau_{2k}={\rm pf}(M_{2k})=\sum_{1\le i_1<\cdots<i_{k}\le n}\left(\prod_{j<l}|z_{i_l}-z_{i_j}|^4\right)E_{i_1}^2\cdots E_{i_{k}}^2\,.
\end{equation}
We emphasize that the form of this $\tau$-function is precisely that of
the partition function of the GSE model.
The fact that the associated eigenvalues appear as doubles is an important
consequence of the self-dual quaternionic structure of the matrices in the
GSE.  

One should compare (\ref{tauGSE}) with the $\tau$-functions of the Toda lattice of $\mathfrak{sl}(n,{\mathbb R})$
which are given by
\[
\tau_{k}^{Toda}=\sum_{1\le i_1<\cdots<i_{k}\le n}\left(\prod_{j<l}|z_{i_l}-z_{i_j}|^2\right)E_{i_1}\cdots E_{i_{k}}\,.
\]
We then expect that the isospectral variety of the Pfaff lattice for GSE model
has a similar
structure as 
that of the Toda lattice, which is that of the permutahedron generated by the
orbit of the symmetric group  $S_n$ (see for example \cite{casian:02, tomei:84}).
The isospectral variety for the GSE-Pfaff lattice will be discussed in Section
\ref{homogeneous}.


\section{Real solutions} \label{realsolutions}
Here we consider generic solutions of the real Pfaff lattice whose initial matrix $L(0)$ has
all real and distinct eigenvalues. 
We start with a brief summary of the results on the skew-orthogonal polynomials obtained in \cite{adler:99, adler:02B} (in particular, we
need Theorem 3.2 in \cite{adler:02B}). The goal of this section is to determine all the fixed points 
of the Pfaff lattice and to find the asymptotic behavior of the solutions.
The geometric structure of the solutions will be
 discussed in the next section.

\subsection{Skew-orthogonal polynomials}
Let us first recall,
\[
\hat L\hat Q=\hat Q L(0)=\hat Q \Psi_0 C_{\gamma}\Psi_0^{-1}\,,
\]
where $L(0)\Phi_0=L(0)\Psi_0V=\Phi_0\Lambda=\Psi_0 C_{\gamma}V$ with $\Psi_0\in \hat{G}_{\mathfrak k}$ and 
the matrix $C_{\gamma}$ is the companion matrix (\ref{companion}). Recall also that the elements on
 the super diagonal of $\hat{L}$ are normalized to all be 1.

The eigenvector $\phi$ of $\hat L$ then defines
skew-orthogonal polynomials as follows:
\[
\hat L\phi=z\phi,\quad \phi=\hat Q \Psi_0\chi\,,
\]
with $\chi=(1,z,\ldots, z^{2n-1})^T$. From $\phi=\hat Q\Psi_0\chi$ with $\hat Q,\Psi_0\in\hat{G}_{\mathfrak k}$, each element of $\phi=(\phi_0,\phi_1,\ldots, \phi_{2n-1})$ is a monic polynomial of
deg$(\phi_k)=k$, and they are given by $\phi_0(z)=1$, $\phi_1(z)=z$ and
\begin{equation}\label{skewpoly}
\begin{cases}
\,\displaystyle{\phi_{2k}(z)}=\displaystyle{z^{2k}+\sum_{j=1}^{2k}q_{2k,2k-j}z^{2k-j},} \\
\,\displaystyle{\phi_{2k+1}(z)}=\displaystyle{z^{2k+1}+\sum_{j=1}^{2k} q_{2k+1,2k-j} z^{2k-j},}
\end{cases}\quad k=1,\ldots,n-1\,,
\end{equation}
where $\{(q_{i,j}) : 0\le i,j\le 2n-1\}$ is the matrix ${\tilde Q}:=\hat
Q\Psi_0$.  
We now see that the equation $\tilde{Q}\tilde{M}\tilde{Q}^T=H^2J$ for the moment matrix
gives the skew-orthogonal relation of the polynomials $\phi_j(z)$:  First define the matrix
$\mathcal{M}: = V^{-1} \tilde{M} V^{-T}$. Then the matrix $\Phi=\tilde{Q}V= (\phi_{i-1}(z_j))_{1\leq i, j \leq 2n}$
satisfies $\Phi\mathcal{M}\Phi^T=H^2J$, which gives  the skew-orthogonal relations,
\[
\langle \phi_{2i},\phi_{2j+1}\rangle_{\mathcal M}=
-\langle \phi_{2j+1},\phi_{2i}\rangle_{\mathcal M}=h_i^2\delta_{i,j}\,,\quad 0\le i,j\le n-1,
\]
and all other cases are zero. Note from $\tilde{M}=V\mathcal{M}V^T$ that the entries $m_{i,j}$ of
$\tilde{M}$ give the moments with respect to the measure $\mathcal{M}$,
\[
m_{i,j}=\langle z^{i-1},z^{j-1}\rangle_{\mathcal M}=\sum_{1\le k,l\le 2n}z_k^{i-1}z_l^{j-1}\mu_{k,l}\,,
\] 
where $\tilde{M}=(m_{i,j})_{1\le i,j\le 2n}$ and $\mathcal{M}=(\mu_{k,l})_{1\le k,l\le 2n}$.

Now solving the skew-orthogonal relation $\Phi\mathcal{M}\Phi^T=H^2J$, we obtain:
\begin{Proposition} {\rm (Theorem 3.1 of \cite{adler:99})} The skew-orthogonal polynomials $\phi_k(z)$ can be
  found by 
\[ \begin{cases}
\,\phi_{2k}(z) = \displaystyle{ \frac{1}{\tau_{2k}(t)} \,
{\rm pf}\begin{pmatrix}
0 & m_{1,2} &  \dots & m_{1,2k+1} & 1 \\
    & 0             & \dots & m_{2,2k+1} & z \\
    &                & \ddots & \vdots      & \vdots \\
    &                &             & 0              & z^{2k} \\
    &                 &            &                  & 0 
\end{pmatrix} \,,}\\
\,\phi_{2k+1}(z) = \displaystyle{ \frac{1}{\tau_{2k}(t)} \,
{\rm pf}\begin{pmatrix}
0 &  m_{1,2}   &     \dots & m_{1,2k}  & 1 & m_{1,2k+2}  \\
    & 0               &     \dots & m_{2,2k}   & z & m_{2,2k+2} \\
    &                  & \ddots   &  \vdots       & \vdots & \vdots \\
    &                  &               & 0                & z^{2k-1} & m_{2k,2k+2}  \\
     &                &                &                   & 0              & -z^{2k+1} \\
      &               &                &                    &                & 0
 \end{pmatrix} \,.}
\end{cases}
\]
for $k=1,2,\ldots, n-1$, with $\phi_0(z)=1$ and $\phi_1(z)=z$.
\end{Proposition} 
To prove this result one notes that  these functions $\phi_j(z)$
satisfy the orthogonality conditions, $\langle \phi_{2k},
z^j \rangle_{\mathcal{M}} = 0\,,\;\mbox{for}\; 0\leq j \leq 2k$, and 
$\langle \phi_{2k+1}, z^j\rangle_{\mathcal M} =0 \,,\; \mbox{for}\; 0\leq j \leq
2k-1\;\mbox{and}\; j=2k+1$.  

The following forms of the skew-orthogonal polynomials are particularly useful:
\begin{Proposition} {\rm (Theorem 3.2 in \cite{adler:02B})} \label{skewtau}
The skew-orthogonal polynomials $\phi_k(z)$ can be expressed in terms of $\tau$-functions,
\[\begin{cases}
\,\phi_{2k}(z)=\displaystyle{\frac{1}{\tau_{2k}(t)}\tau_{2k}\left(t-\left[z^{-1}\right]\right)
z^{2k}, }\\
\,\phi_{2k+1}(z)=\displaystyle{\frac{1}{\tau_{2k}(t)}\left(z+\frac{\partial}{\partial t_1}\right)
\tau_{2k}\left(t-\left[z^{-1}\right]\right) z^{2k}, }
\end{cases}
\]
where $\tau_{2k}(t-[{z^{-1}}])=\tau_{2k}(t_1-\frac{1}{z},t_2-\frac{1}{2z^2},\ldots)$.
\end{Proposition}
\noindent
(A proof of this Proposition is based on the moment equation given in (\ref{momentequation}), and
Lemma \ref{Stau} below.)

  From Proposition \ref{skewtau}, we have the explicit form of the entries $q_{i,j}$ of $\tilde Q$  in (\ref{skewpoly}) in terms of the $\tau$-functions,
 \begin{equation}\label{tildeQ}\begin{cases}
\,q_{2k,2k-j}(t)=\displaystyle{\frac{S_j(-\tilde\partial)\tau_{2k}(t)}{\tau_{2k}(t)}, }\\
\,q_{2k+1,2k-j}(t)=\displaystyle{\frac{[S_{j+1}(-\tilde\partial)+\partial_1S_j(-\tilde\partial)]\tau_{2k}(t)}{\tau_{2k}(t)}}
\, .
\end{cases}
\end{equation}
Here $S_k(-\tilde\partial)$ denotes $S_k(-\partial_1,-\frac{1}{2}\partial_2,-\frac{1}{3}\partial_3,\ldots)$,
with $\partial_k=\frac{\partial}{\partial t_k}$, and $S_k(t_1,t_2,\ldots)$ are symmetric functions defined by
\begin{equation}\label{elementary}
\exp\left(\sum_{j=1}^{\infty}t_jz^j\right)=\sum_{k=0}^{\infty}S_k(t_1,\ldots,t_k)z^k\,.
\end{equation}
The explicit form of the $S_k(t_1,\ldots,t_k)$ is given by
\[
S_k(t_1,t_2,\ldots,t_k)=\sum_{j_1+2j_2+\cdots+kj_k=k}\, \frac{t_1^{j_1}t_2^{j_2}\cdots t_k^{j_k}}{j_1!
j_2! \cdots j_k!}\,,
\]
which is the complete homogeneous symmetric function $h_k(x_1,\ldots,x_{2n-1})$ with
$t_k=\frac{1}{k}\sum_{i=1}^{2n-1}x_i^{k}$ (see \cite{macdonald:79}),
\[
S_k(t_1,\ldots,t_k)=h_k(x_1,\ldots,x_{2n-1})=\sum_{i_1+i_2+\cdots +i_{2n-1}=k}x_1^{i_1}x_2^{i_2}\cdots x_{2n-1}^{i_{2n-1}}\,.
\]
With the $\tilde Q$ given in (\ref{tildeQ}), we can find $\hat L$ in terms of the $\tau$-functions,
\begin{equation}\label{Lhat}
\hat L=\tilde Q C_{\gamma}\tilde Q^{-1}\,.
\end{equation}
In particular, the $2\times 2$ block diagonal part of $\hat L$ has the form,
 $\hat L_0:={\rm diag}_2(\hat L_{0,0},\hat L_{1,1},\ldots, \hat L_{n-1,n-1})$ with
\begin{equation}\label{L0}
\hat L_{k,k}=\begin{pmatrix}
q_{2k,2k-1}  &   1   \\
q_{2k+1,2k-1}-q_{2k+2,2k} & - q_{2k+2,2k+1}
\end{pmatrix} \quad k=1,2,\ldots, n-2,
\end{equation}
and
\[
\hat L_{0,0}=\begin{pmatrix}
0  &   1\\
-q_{2,0} & -q_{2,1}
\end{pmatrix}\,,\quad \quad
\hat L_{n-1,n-1}=\begin{pmatrix}
q_{2n-2,2n-3}  &   1   \\
q_{2n-1,2n-3}-\gamma_2 & 0
\end{pmatrix}\,.
\]
Notice from (\ref{tildeQ}) that the diagonal elements of $\hat L$ (also for $L$ in (\ref{L})) are
given by $b_k=-q_{2k,2k-1}=\frac{\partial}{\partial t_1}\ln\tau_{2k}$. The formulae
for $b_k$ together with $a_k$ in (\ref{a}) are
similar to the formulae for the dependent variables in the Toda lattice, i.e.
\begin{equation}\label{ab}
a_k=\frac{\sqrt{\tau_{2k-2}\tau_{2k+2}}}{\tau_{2k}},\quad b_k=\frac{\partial}{\partial t_1}\ln\tau_{2k}\,,
\qquad {\rm for} \quad k=1,\ldots,n-1.
\end{equation}
In particular, the functions $b_k(t)$ determine the diagonal elements of $\hat{L}$, and
provide the basic information about the geometry of the isospectral variety, similar to the
Schur-Horn theorem for the set of symmetric matrices (see Section \ref{homogeneous}).


\subsection{Fixed points of the Pfaff flows}

We now compute the asymptotic values of $\hat L$ for the case of $L(0)$ whose eigenvalues are
all real and to be ordered as
\[
z_1<z_2<\cdots <z_{2n}\,.
\]
In the case of GSE-Pfaff lattice, we take the limit $z_{2k}\to z_{2k-1}$.

Let us start with the following Lemma for computing
 $S_j(-\tilde\partial)\tau_{2k}$:
\begin{Lemma}\label{Stau}
With $E_k(t)=\exp\theta(t,z_k)$, we have, for $m\ge j$,
\[
S_j(-\tilde\partial)\,E_1\cdots E_m=(-1)^j\sigma_j(z_1,\ldots,z_m)\,E_1\cdots E_m\,,
\]
where $\sigma_j(z_1\ldots,z_m)$ is the elementary symmetric polynomial of degree $j$ in $(z_1,\ldots,z_m)$,
\[
\sigma_j(z_1,\ldots,z_m)=\sum_{1\le i_1<\cdots<i_j\le m}z_{i_1}\cdots z_{i_j}\,.
\]
(note $\gamma_k=\sigma_k(z_1,\ldots,z_{2n})$ in $C_{\gamma}$). 
For $j>m$, we have $S_j(-\tilde\partial)E_1\cdots E_m=0$.
\end{Lemma}
\begin{Proof}
Since ${\frac{\partial}{\partial t_j}E_k=\frac{\partial^j}{\partial t_1^j}E_k}$, we have
\begin{align*}
E_k(t-[z^{-1}])&=\displaystyle{\exp\left(-\sum_{j=1}^{\infty}\frac{1}{jz^j}\frac{\partial}{\partial t_j}\right) E_k(t)=\exp\left(-\sum_{j=1}^{\infty}\frac{1}{jz^j}\frac{\partial^j}{\partial t_1^j}\right)E_k(t)}\\
    &=\displaystyle{\exp\left[\ln\left(1-\frac{1}{z}\frac{\partial}{\partial t_1}\right)\right]E_k(t)=
    \left(1-\frac{z_k}{z}\right)E_k(t)}\,.
\end{align*}
Then from (\ref{elementary})
we find
\begin{align*}
\displaystyle{\sum_{j=0}^{\infty}\frac{1}{z^j}S_j(-\tilde\partial)\,(E_1\cdots E_m)(t)}&=
(E_1\cdots E_m)(t-[z^{-1}]) \\
&=\displaystyle{\prod_{i=1}^m\left(1-\frac{z_i}{z}\right)\, (E_1\cdots
  E_m)(t)} \,.
\end{align*}
Comparing the coefficients of $z^{-k}$, we obtain the desired formulae.
\end{Proof}
This lemma confirms again that $\phi_{2n}(z) = \mbox{det}(L-zI)$.  
 From this Lemma and the ordering of the $z_k$'s, we have, for $t_1\to\infty$,
\begin{equation}\label{asymptoticQ}\begin{cases}
\,{q_{2k,2k-j}(t)\longrightarrow (-1)^j\sigma_j({2k})},\\
\,{q_{2k+1,2k-j}(t)\longrightarrow (-1)^{j+1}\sigma_{j+1}({2k})+(-1)^j\sigma_1(2k)\sigma_{j}({2k})}\,.
\end{cases}
\end{equation}
Here we introduced the notation $\sigma_1(2k):=\sigma_1(z_1,\ldots,z_{2k})$.
We then have
\begin{Theorem}\label{asymptoticL}
In generic case, the $\hat L(t)$ approaches the block upper triangular
matrix $\hat L^{-}$ as $t_1\to -\infty$,
\[
\hat L^{-}=
\begin{pmatrix}
\hat L_{0,0}^{-} &  e_-& 0_2 &\cdots &0_2\\
 0_2    & \hat L_{1,1}^{-} & e_-& \cdots &0_2\\
\vdots    &  \vdots   & \ddots &\ddots &\vdots\\
0_2 & 0_2 &\cdots &\hat L^{-}_{n-2,n-2} &e_-\\
0_2  &  0_2  &  \cdots &\cdots & \hat L^{-}_{n-1,n-1}
\end{pmatrix} \,\in {\mathfrak g}_0\oplus{\mathfrak g}_+\,,
\]
where $e_-=\begin{pmatrix}0&0\\1&0\end{pmatrix}$ and $\hat L_{k,k}^{-}$ is the $2\times 2$
matrix defined by
\[
\hat L^{-}_{k,k}=\begin{pmatrix}
-\sigma_1({2k})  &  1   \\
-\left(\sigma_1({2k})+z_{2k+1}\right)\left(\sigma_1({2k})+z_{2k+2}\right) & \sigma_1({2k+2})
\end{pmatrix}\quad k=0,1,\ldots,n-1,
\]
with $\sigma_1(m)=\sigma_1(z_1,\ldots,z_m)=\sum_{j=1}^mz_j$ and $\sigma_1(2n)={\rm tr}(\hat L)=0$.

On the other hand, as $t_1\to \infty$, the $\hat L(t)$ approaches  $\hat L^+$ whose $2\times 2$ block diagonal elements are given by $\hat L^{+}_{k,k}=\hat L^{-}_{n-k-1,n-k-1}$ for $k=0,1,\ldots,n-1$.
\end{Theorem}
\begin{Proof}
First note from (\ref{tildeQ}) that $\tilde Q = \hat{Q}\Psi_0$ approaches a constant matrix as $t_1\to\pm\infty$. This implies $\hat Q^{-1}\frac{\partial \hat Q}{\partial t_1}\to 0$ as $t_1\to\pm\infty$.
On the other hand, from Lemma \ref{LemmaQ}, we have
\[
\hat B_1=H\left(L_--JL_+^TJ\right)H^{-1}=\hat L_--JHL_+^TH^{-1}J\,.
\]
The non zero entries in $H(L_+)^TH^{-1}$ are given by $a_k^2$ (see (\ref{a})). With the ordering
of the eigenvalues, i.e. $z_1<z_2<\cdots<z_{2n}$, and the genericity, one can see
\[
\tau_{2k}(t)\longrightarrow\begin{cases} c^- E_1\cdots E_{2k},  &\text{as $ t_1\to-\infty $;}\\
c^+E_{2n-2k+1}\cdots E_{2n},&\text{ as $ t_1\to\infty$,}
\end{cases}
\]
where $c^{\pm}$ are nonzero constants.
Then from (\ref{a}), we have $a_k(t)\to 0$ as $t_1\to\pm\infty$, therefor
$L_+\to 0$.  Since
$\hat{B}_1 \to 0$ and $J H L_+^T H^{-1} J \to 0$ we conclude that $\hat{L}_-
\to 0$.  

  Since $\hat L_+$ is just a constant matrix, 
it only remains to find the limits of $\hat L_0$ as $t\to \pm \infty$.
 The $\hat L_0$ is explicitly
  given by (\ref{L0}), and using (\ref{asymptoticQ}), we can easily compute $\hat L_0$ for $t_1\to\pm\infty$.
\end{Proof}

Each submatrix $\hat L_{k,k}^{-}$ in the asymptotic value of $\hat L$ has the eigenvalues
$\{z_{2k+1},z_{2k+2}\}$. This implies that the Pfaff lattice has the sorting property of eigenvalues in
 pairs as $t_1\to\pm\infty$, i.e. $(\{z_1,z_2\},\ldots, \{z_{2n-1},z_{2n}\})$ as $t_1\to-\infty$
and $(\{z_{2n-1},z_{2n}\},\ldots,\{z_1,z_2\})$ as $t_1\to \infty$. It should be noted that
the order in the pair is free.

We can also show:
\begin{Lemma}\label{fixedPt}
A matrix $\hat{L}\in\hat{Z}_{\mathbb R}(\gamma)$ is a fixed point of the Pfaff lattice,
if and only if 
 $\hat{L}(t)$ is a block upper triangular
matrix for all $t$.
\end{Lemma}
\begin{Proof}
Let us first note that the Pfaff lattice
equation for $\hat{L}$ is given by
\[
\frac{\partial\hat{L}}{\partial t_j} =-\left[\hat{B}_j,\hat{L}\right]\,,
\]
where $\hat{B}_j$ is given in Lemma \ref{LemmaQ}. In particular, we
have $\hat{B}_1=\hat{L}_--JHL_+^TH^{-1}J$, and with $a_k=h_{k+1}/h_k$, the second term can be
explicitly expressed as
\[
JHL_+^TH^{-1}J=JH^2\hat{L}_+^TH^{-2}J=\sum_{k=1}^{n-1}a_k^2\,e_{2(k+1),2k-1}\,,
\]
where $e_{i,j}$ is the $2n\times 2n$ matrix with $1$ at $(i,j)$-entry
and $0$ for all others. Then if $\hat{L}$ is block upper triangular, i.e. $\hat{L}_-=0$,
we can compute directly that $[\hat{B}_1, \hat{L}]$ is
strictly block lower triangular with entries given only by the $a_k$'s.  This implies 
$ \frac{\partial\hat{L}}{\partial t_1}
= 0$, and therefore $[\hat{B}_1, \hat{L}] =0$ which
fixes $a_k=0$ for all $k$.  One also notes that if $\hat{L}(t)$ is not
block upper triangular, then $\frac{\partial\hat{L}}{\partial t_1} \neq 0$, therefore the
only possible fixed points are those of block upper triangular shape.
A similar direct calculation gives the
same conclusions for the higher flows with generator $\hat{B}_j$ for $j\ge 2$.
\end{Proof}
We can also obtain the explicit block upper triangular 
form of a fixed point, $\hat L \in
\hat Z_{\mathbb{R}}(\gamma)$, of the Pfaff flow; we find 
that a general fixed point is in a form 
similar to that of $\hat L^-$ in Theorem \ref{asymptoticL}:
\begin{Theorem}\label{fixedL}
Each fixed point of the Pfaff lattice is uniquely parametrized by a set
\[ \label{set_of_pairs}
(\{z_{i_{1}},z_{i_2}\},\ldots,\{z_{i_{2n-1}},z_{i_{2n}}\}):\;\mbox{with}\;\;
 i_{2k-1}<i_{2k},\; \;  k=1,\ldots,n\,.
\]
Therefore the total number of fixed points is given by $(2n)!/2^n$.
Moreover the fixed point corresponding to this set is expressed by a block upper triangular matrix in $\hat{Z}_{\mathbb R}(\gamma)$
whose block diagonal is given by ${\rm diag}_2(\hat{L})=\left(\hat{L}_{0,0},\hat{L}_{1,1},\ldots,\hat{L}_{n-1,n-1}\right)$
with
\[
\hat L_{k,k}=\begin{pmatrix}
-\sigma_1(i_1,\ldots,i_{2k})      &    1   \\
-(\sigma_1(i_1,\ldots,i_{2k})+z_{i_{2k+1}})(\sigma_1(i_1,\ldots,i_{2k})+z_{i_{2k+2}}) & \sigma_1(i_1,\ldots,i_{2k+2})\end{pmatrix}\,.
\]
where $\sigma_1(i_1,\ldots,i_{m})=\sum_{j=1}^{m}z_{i_j}$ and $\sigma_1(0)=0$.
\end{Theorem}
\begin{Proof}
For a block upper triangular matrix
with the eigenvalues $\{z_1,\ldots, z_{2n}\}$,
each $2\times 2$ matrix in the $2\times 2$
block diagonal part  has a pair of eigenvalues $\{z_{i_{2k-1}},z_{i_{2k}}\}$
with $1\le i_{2k-1}<i_{2k}\le 2n$, therefore a fixed point of
the Pfaff lattice gives the set above.
Then from the form of $\hat{L}$ in $\hat{Z}_{\mathbb R}(\gamma)$, one can construct a unique $\hat{L}$ corresponding to this set: In the first block (the top left $2\times 2$ submatrix $\hat{L}_{0,0}$), we have a unique block,
\[
\begin{pmatrix}0 & 1\\-z_{1_1}z_{i_2}&z_{i_1}+z_{i_2}\end{pmatrix}\,.
\]
Now suppose $\hat{L}_{k,k}$ has the above form, then, as $\hat L \in
\hat Z_\mathbb{R}(\gamma)$,  $\hat{L}_{k+1,k+1}$ has the form
\[
\begin{pmatrix}
-\sigma_1(i_1,\ldots,i_{2k+2}) & 1 \\  x  &  b_{k+2} \end{pmatrix}\,.
\]
Since this block has the eigenvalues $\{z_{i_{2k+3}},z_{i_{2k+4}}\}$, one finds the unique 
$b_{k+2}=-\sigma_1(i_1,\ldots,i_{2k+4})$ and $x=-(\sigma_1(i_1,\ldots, i_{2k+2})+z_{i_{2k+3}})(\sigma_1(i_1,\ldots,i_{2k+2})+z_{i_{2k+4}})$. This completes the proof.
\end{Proof}
In the next section, we identify those fixed points as the vertices of the moment polytope
generated by the flows associated with the Chevalley invariants, that is, the torus-fixed points
of the Pfaff lattice.

\begin{figure}[t]
\includegraphics[width=13.5cm]{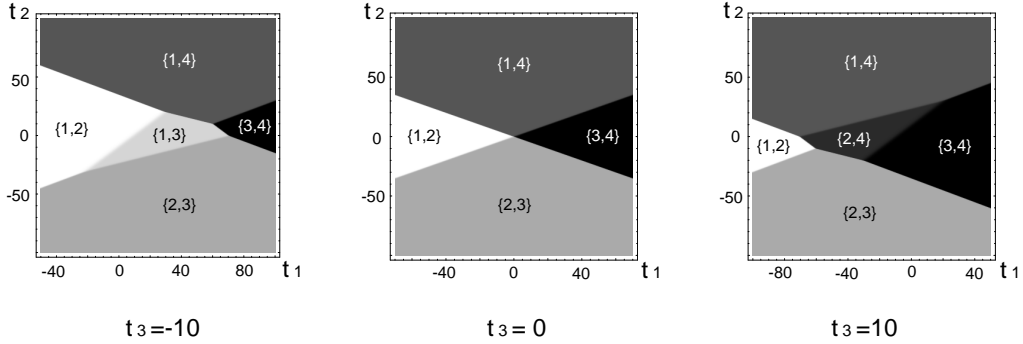}
\caption{The evolution of  $b_1(t_1,t_2,t_3)$, the diagonal element of $L$ for the GOE-Pfaff lattice
with $n=2$. The pair of numbers $\{i,j\}$ indicates
the value of $b_1$ in the corresponding region, i.e. $b_1\approx z_i+z_j$.
Notice that the pairs $\{1,3\}$ and $\{2,4\}$ appear in the interaction region forming a double
quadrangle cone in the $t_3$-direction with a singular point at $t_1=t_2=t_3=0$.}
\label{2sol:fig}
\end{figure}

\begin{Example}\label{n2example}
The case $n=2$ for the GOE-Pfaff lattice: The $\tau_2$-function is given by
\[
\tau_2=\sum_{1\le i<j\le 4}b_{i,j}E(i,j)\,.
\]
where $E(i,j)=(z_j-z_i)E_iE_j$ with the ordering $z_1<z_2<z_3<z_4$.
 We have $4!/2^2=6$ fixed points, and each fixed point corresponds to the exponential 
term $b_{i,j}E(i,j)$ for $1\le i<j\le 4$.
We consider the Pfaff orbit in the space of $(t_1,t_2,t_3)$ corresponding to the
Chevalley invariants, $H_k(L):=\frac{1}{k+1}{\rm tr}(L^{k+1})$ for $k=1,2,3$,
that is, we consider (\ref{lax}).  We consider the asymptotic values of the diagonal
element of $L(t)$, i.e.
\[
b_1(t)=\frac{\partial}{\partial t_1}\ln\,\tau_2(t)\,.
\]
If the exponential term $E(i,j)$ in the $\tau_2$ dominates the others in some region
of $(t_1,t_2,t_3)$, then in this region $b_1$ takes the value
\[
b_1(t_1,t_2,t_3)\approx z_i+z_j\,. 
\]
One can then identify each fixed point with the ordered set of pairs $(\{z_i,z_j\},\{z_k,z_l\}):\;\mbox{for}\; i<j,\;\mbox{and}\; k<l$ as the region where 
$E(i,j)$ gives the dominant term. So we expect to have six different regions in $(t_1,t_2,t_3)$ space:
Figure \ref{2sol:fig} plots $b_1(t)$ for the case of $n=2$ in the GOE setting where
\[
B=\begin{pmatrix}
0 & 1 & 1 & 1\\
  &  0 &  1  & 1\\
   &    &   0  &  1  \\
   &     &       &   0
\end{pmatrix}\,,
\]
with $(z_1,z_2,z_3,z_4)=(-2,-1,0,3)$, that is, we have
\[
\tau_2(t)=E_1E_2+2E_1E_3+5E_1E_4+E_2E_3+4E_2E_4 +3E_3E_4\,.
\]
In Figure \ref{2sol:fig}, each region with the dominant $E(i,j)$ is marked by $\{i,j\}$,
and the boundaries of the regions are given by the balance of two exponential terms,
e.g. $E_1E_2=2E_1E_3$ which gives $\theta(t,z_2)=\theta(t,z_3)+\ln 2$. This equation
describes the boundary line between $\{1,2\}$ and $\{1,3\}$ in the $t_1$-$t_2$ plane 
for each fixed $t_3$,
\[
(z_2-z_1)t_1+(z_2^2-z_1^2)t_2+(z_2^3-z_1^3)t_3={\rm constant}.
\]
This line expresses a line soliton solution of the coupled KP (or DKP) equation,
and Figure \ref{2sol:fig} illustrates a resonant interaction of 2-soliton solution
(see \cite{kodama:06}).
\end{Example}





\subsection{The GSE-Pfaff lattice}
Recall that the moment matrix for the GSE model was obtained by taking the limits $z_{2k}\to z_{2k-1}$,
and setting $b_{2k-1,2k}=\frac{c_k}{z_{2k}-z_{2k-1}}$ for some constants $c_k$ and
 $b_{i,j}=0$ for all others. Note here that if $c_k>0$, then it can be
 absorbed into the exponential term $E_{2k-1} E_{2k}$.
 However those constants will be important for the general solutions including
 rational solutions of the Pfaff lattice (see below).
We also relabel the eigenvalues $z_{2k-1}$ as $z_k$  for convenience as in
Section \ref{GSEmatrixmodel}.
Then we have
\begin{Theorem}\label{GSEasymptotics}
As $t_1\to-\infty$, the $\hat L(t)$ for the GSE-Pfaff lattice approaches the upper triangular
matrix $\hat L^-$, as in Theorem \ref{asymptoticL}, whose block diagonal
elements are now given by
\[
\hat L^-_{k,k}=\begin{pmatrix}
-2\sigma_1(k)   &   1  \\
-(2\sigma_1(k)+z_{k+1})^2  &  2\sigma_1(k+1)
\end{pmatrix}\quad\quad k=0,1,\ldots,n-1,
\]
with $\sigma_1(k)=\sum_{j=1}^kz_j$, and $\sigma_1(0) = \sigma_1(n) = 0$.
\end{Theorem}
\begin{Proof}
In the limit $z_{2k}\to z_{2k-1}$, the $\sigma_1(2k)$ becomes $2\sum_{j=1}^kz_{2j-1}$,
which is just $2\sigma_1(k)$ after relabeling the eigenvalues.
\end{Proof}
Notice that the matrix $\hat L^-_{k,k}$ has doubly degenerate eigenvalues
$z_{k+1}$.  Proposition \ref{GSEasymptotics}  means that the GSE-Pfaff
lattice flow sorts the double eigenvalues in the same manner that the Toda
lattice flow sorts eigenvalues.    
One can also see that there are $n!$ number of fixed points, and
each fixed point corresponds to the matrix having the permuted double eigenvalues,
\[
(z_{i_1},z_{i_2},\ldots, z_{i_n})\,.
\]
Those points form the vertices of the permutahedron of the orbit of the permutation
group, which is the isospectral manifold of the Toda lattice (see
e.g. \cite{tomei:84, casian:02}, also see Section \ref{GSEpolytope}).

\begin{figure}[t]
\includegraphics[width=9.5cm]{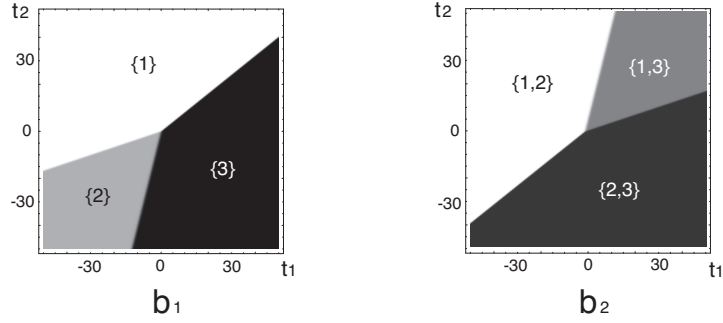}
\caption{The graphs of the diagonal elements $b_1$ and $b_2$ of $L$ for the GSE-Pfaff lattice
with $n=3$. The number sets $\{i\}$ and $\{j,k\}$ with $j<k$ indicate
the values of $b_{1}$ and $b_2$, i.e. $b_1=2z_i$ and $b_2=2(z_j+z_k)$ in those
regions.}\label{gse:fig}
\end{figure}

\begin{Example} The GSE-Pfaff lattice with $n=3$: The $\tau$-functions are given by
\begin{align*}
\tau_2&=c_1E_1^2+c_2E_2^2+c_3E_3^2\,,\\
\tau_4&=c_1c_2(z_1-z_2)^4E_1^2E_2^2+c_1c_3(z_1-z_3)^4E_1^2E_3^2+c_2c_3(z_2-z_3)^4E_2^2E_3^2\,,
\end{align*}
where $c_k$ are nonnegative constants, and $E_k=e^{\theta(t,z_k)}$ with $\theta(t,z_k)=\sum_{j=1}^3 t_jz_k^j$. Each exponential term
in $\tau_{2k}$ becomes dominant in a certain region of the $(t_1,t_2,t_3)$ space.
Figure \ref{gse:fig} plots the functions $b_k=\frac{\partial}{\partial t_1}\ln\tau_{2k}$ for $k=1,2$
to show the dominant exponentials in the $\tau$-functions.  Then
we obtain the fixed points of the Pfaff lattice. For example, the region having
$b_1=2z_2=\{2\}$ and $b_2=2(z_1+z_2)=\{1,2\}$ gives the fixed point
$\hat L^-={\rm diag}_2(\hat L^-_{0,0},\hat L^-_{1,1}.\hat L^-_{2,2})$ with
\[
\hat L_{0,0}^-=\begin{pmatrix} 0 & 1\\ -z_2^2 &2z_2\end{pmatrix}\,,\quad
\hat L_{1,1}^-=\begin{pmatrix} -2z_2 & 1\\ -(z_2-z_3)^2 & -2z_3\end{pmatrix}\,,\quad
\hat L_{2,2}^-=\begin{pmatrix} 2z_3 & 1\\ -z_3^2 & 0 \end{pmatrix}\,,
\]
where we have used $z_1+z_2+z_3=0$. This fixed point corresponds to the ordered
eigenvalues $(z_2,z_1,z_3)$.
\end{Example}

\begin{Remark} The rational solutions of the Pfaff lattice discussed in \cite{adler:02}
are given by the GOE $\tau$-functions in the nilpotent limit with specialized choices of
the $B$-matrix. (By the nilpotent limit, we mean the limit $z_k\to 0$ for all $k$.): We compute the moment matrix
$\tilde M=EBE^T=e^{\theta(t,C_{\gamma})}\tilde B e^{\theta(t,C_{\gamma})^T}$ with
 $\tilde B=VBV^T$ in the nilpotent limit corresponding to $\gamma=0$ (i.e. $z_k=0$ for all $k$). Choosing an appropriate
 $B$-matrix so that there exists a limit $\tilde B_0=\lim_{\gamma\to 0}VBV^T$, we have
$\tilde M_0=e^{\theta(t,C_0)}\tilde B_0 e^{\theta(t,C_0)^T}$ with
\[
e^{\theta(t,C_0)}=\begin{pmatrix} 
1 & S_1(t) & S_2(t) & \cdots  & S_{2n-1}(t) \\
0 &   1       &  S_1(t) & \cdots  & S_{2n-2}(t) \\
0 & 0       &  1     & \cdots    &     S_{2n-3}(t)  \\
\vdots& \vdots   &\vdots  &\ddots  &\vdots  \\
0    &   0     &    0  &\cdots  &   1
\end{pmatrix}\,,
\]
where $S_k(t)=S_k(t_1,\ldots,t_k)$ are the complete homogeneous symmetric functions defined in
(\ref{elementary}).
Various choices of $\tilde B_0$ are considered in \cite{adler:02},
some of which contain the Jack polynomials. In particular, if $\tilde{B}_0$ is chosen as 
the matrix having only nonzero anti-diagonal elements with $b_{k,2n-k+1}=2(n-k)+1$ for
$k=1,\ldots,n$, the Pfaffian $\tau_{2k}={\rm pf}((\tilde{M}_0)_{2k})$ then
gives the Jack polynomial $J^{(1/2)}_{\lambda}(x)$ for rectangular partitions, 
$\lambda=(\,\overbrace{2(n-k),\ldots,2(n-k)}^k\,)$ with $t_j=\frac{1}{j}\sum_i x_i^j$,
which is a zonal spherical function on $GL(2n)/Sp(n)$. It is then interesting to compare this with the case of the nilpotent Toda lattice
of ${\mathfrak{sl}}(n)$ where
the $\tau$-functions are given by the Schur polynomials with rectangular partitions,
i.e. $\tau_k^{Toda}=(-1)^{\frac{k(k-1)}{2}}S_{\lambda}(t_1,\ldots,t_{n-1})$ with
$\lambda=(\,\overbrace{n-k,\ldots,n-k}^k\,)$ for $k=1,\ldots,n-1$ (see \cite{casian:04}).

\end{Remark}


\section{Geometry of the isospectral variety}\label{homogeneous}
Here we discuss the geometric structure of the isospectral variety for the Pfaff lattice using
the moment map defined in \cite{gelfand:87}.

\subsection{Geometric structure of the $\tau$-functions}
Recall that the solution of the Pfaff lattice is given by the matrix factorization, $e^{\theta(t,L(0))}=Q(t)^{-1}P(t)$
and $L(t)=Q(t)L(0)Q(t)^{-1}$.
This implies that we have a bijection between
the isospectral variety $Z_{\mathbb R}(\gamma)$ in (\ref{isospectralZ}) and the homogeneous space,
\[
{\mathcal H}:=\frac{SL(2n,{\mathbb R})}{Sp(n,{\mathbb R})}\,.
\]
with dim$\,{\mathcal H}=2n^2-n-1$. Note that ${\mathcal H}$ is a symmetric space associated
with the involution $\theta(\exp{X})=\exp(\sigma(X))$ on $SL(2n)$ with $\sigma(X)=JX^TJ$,
i.e. $\sigma=d\theta$. 
Then each point of the Pfaff orbit can be parametrized by a skew-symmetric matrix.
The $B$-matrix used in the $\tau$-functions
indeed gives a parametrization of the point of ${\mathcal H}$. Namely we have
\[
\begin{array}{ccccccc}
Z_\mathbb{R}(\gamma) &\overset{c_{\gamma}}{\longrightarrow} &{\mathcal H} &\overset{\phi}\longrightarrow & {Skew}(2n) \\
&{}& \\
L(0) &\longmapsto & \Psi_0^{-1} \mod Sp(n,{\mathbb R}) &\longmapsto & \tilde B=\Psi_0^{-1}J\Psi_0^{-T}\\
\end{array}
\]
where $\tilde B=VBV^T$ and $L(0)\Psi_0=\Psi_0C_{\gamma}$ with $\Psi_0\in\hat {G}_{\mathfrak k}$. Here the $\phi$ gives an isomorphism when $\phi$ is restricted on
an open dense subset of ${\mathcal H}$, and ${Skew}(2n)$ is the set of skew-symmetric
matrices with det$\,=1$. The $B$-matrix then parametrizes an initial point of the Pfaff flow, and the time evolution follows
\[\begin{array}{ccccccc}
L(t) &\longmapsto & e^{\theta(t,C_{\gamma})}\Psi_0^{-1} \mod Sp(n,{\mathbb R}) &\longmapsto & 
\tilde M=E(t,\Lambda)BE(t,\Lambda)^T\end{array}
\]
Note here that $\tilde M=e^{\theta(t,C_{\gamma})}\tilde{B} e^{\theta(t,C_{\gamma})^T}$ with $E(t,\Lambda)=Ve^{\theta(t,\Lambda)}=e^{\theta(t,C_{\gamma})}V$. The Pfaff flow also defines a torus action
on ${\mathcal H}$, that is, $e^{\theta(t,C_{\gamma})}\Psi_0^{-1}Sp(n)=Ve^{\theta(t,\Lambda)}\Phi_0^{-1}Sp(n)$.

Each $\tau$-function $\tau_{2k}$ can be expressed in the following form, which is a Pfaffian
version of the Binet-Cauchy theorem \cite{ishikawa:95}:
\begin{Lemma} \label{dtauBC} 
\[
\tau_{2k}={\rm pf}(\tilde{M}_{2k})=\sum_{1\le i_1<\cdots<i_{2k}\le 2n} {\rm pf}(i_1,\ldots,i_{2k}) {E}(i_1,\ldots,i_{2k})\,.
\]
where
\begin{itemize}
\item{} ${E}(i_1,\ldots,i_{2k})$ is the determinant of the $2k\times 2k$ submatrix of the Wronskian matrix $E(t,\Lambda)$ in (\ref{wronskianE}), and is  given by
\[
{E}(i_1,\ldots,i_{2k})={\rm det}\begin{pmatrix}
E_{i_1} & E_{i_2} & \cdots  & E_{i_{2k}} \\
 E'_{i_1} &   E'_{i_2}    & \cdots   &E'_{i_{2k}}  \\
 \vdots &               &  \ddots & \vdots \\
  E^{(2k-1)}_{i_1}&  E^{(2k-1)}_{i_2}     & \cdots& E^{(2k-1)}_{i_{2k}} 
  \end{pmatrix}\,.
  \]

\item{} ${\rm pf}(i_1,\ldots,i_{2k})$ is the Pfaffian of the $2k\times 2k$ skewsymmetric submatrix of the $B$-matrix, and is
defined by
\[
{\rm{pf}}(i_1,\ldots,i_{2k})={\rm pf}\begin{pmatrix}
0 & b_{i_1,i_2} & \cdots & \cdots & b_{i_1,i_{2k}} \\
 &   0    & \cdots   &  \cdots &b_{i_2,i_{2k}}  \\
 &        &                 &  \cdots & \vdots \\
  &       &     & 0 & b_{i_{2k-1},i_{2k}}  \\
  &        &     &        &    0
  \end{pmatrix}\,.
  \]
\end{itemize} 
\end{Lemma}
Recall that $\{ E_i: i=1, \ldots, 2n\}$ forms a basis of
$\mathbb{R}^{2n}$ for distinct $z_k$, let us further assume that
$z_{i_1, \ldots, i_{2k}} = z_{i_1} + \cdots + z_{i_{2k}}$ are all
distinct.  
There is a bijection between $E(i_1, \ldots, i_{2k})$ and the basis
vector $e_{i_1} \wedge e_{i_2} \wedge \dots \wedge e_{i_{2k}}$ of
$\wedge^{2k}{\mathbb{R}^{2n}}$, but this bijection is not a homomorphism of
  the $\wedge$-product structure on the space $\bigoplus_{k=1}^n
  \wedge^{2k} \mathbb{R}^{2n}$, this deficiency does not affect our
  computation.   
 The coefficient of the
$E(i_1,\ldots,i_{2k})$  gives a Pfaffian-Pl\"ucker coordinate,
which is written recursively by the expansion formula,
\[
{\rm pf}(i_1,i_2,\ldots,i_{2k})=\sum_{j=2}^{2k}(-1)^jb_{i_1,i_j}\,{\rm pf}(i_2,\ldots,\hat i_j,\ldots,i_{2k})\,,
\]
where $\hat i_j$ implies the deletion of $i_j$ from
pf$(i_1,\ldots,i_{2k})$, and ${\rm pf}(i, j) = b_{i,j}$.
For example, 
the Pfaffian pf$(1, 2, 3, 4)$ is given by 
\[
{\rm pf}(1, 2 , 3, 4)=b_{1,2}b_{3,4}-b_{1,3}b_{2,4}+b_{1,4}b_{2,3}\,.
\]
One should also note from (\ref{tildeQ}) that each element $q_{r,k}$ in $\tilde{Q}$ is invariant under the
scalar multiplication to $\tau_{2k}$ as is $\hat{L}$ from
(\ref{Lhat}),  that is, $\tau_{2k}\equiv c_k\tau_{2k}$ for any
constant $c_k$. This means that the $\tau_{2k}$ can be considered as a
point of the projective space ${\mathbb P}(\wedge^{2k}{\mathbb R}^{2n})$. 

We note here that $\tau_{2k}$ is not (in general) decomposable in
$\wedge^{2k} \mathbb{R}^{2n}$, therefore $\tau_{2k}$ does not
represent a point in the Grassmannian $Gr(2k, 2n)$.

\subsection{Moment polytope}
Let us recall that the unipotent subgroup $\hat {G}_{\mathfrak k}$ can be identified as
the top cell ${\Sigma}_{id}$  of the flag variety $SL(2n,{\mathbb R})/P_{2\times 2}$ (see Section \ref{moment-tau}).
It is also well-known that there is an embedding of the flag $SL(2n,{\mathbb R})/P_{2\times 2}$,
\[
SL(2n,{\mathbb R})/P_{2\times 2}~\hookrightarrow~{\mathbb P}(\wedge^2{\mathbb R}^{2n})\times{\mathbb P}(\wedge^4{\mathbb R}^{2n})\times\cdots
\times{\mathbb P}(\wedge^{2n-2}{\mathbb R}^{2n})\,.
\]
This suggests that $\hat{G}_\mathfrak{k}$ can be embedded in this
product space (see (\ref{immersion})).

 On each ${\mathbb P}(\wedge^{2k}{\mathbb R}^{2n})$, one can define a moment map,
$ \mu : {\mathbb P}(\wedge^{2k}{\mathbb R}^{2n})  \longrightarrow  {\mathfrak h}_{\mathbb R}^* $, with
a torus action given by the Pfaff flow $g=e^{\theta(t,\Lambda)}={\rm diag}(E_1,\ldots,E_{2n})$
(see \cite{gelfand:87, shipman:00}): By considering $\tau_{2k}$ to
be a point on ${\mathbb P}(\wedge^{2k}{\mathbb R}^{2n})$, we define
the image $\mu(\tau_{2k})$ as follows:  For the expression of
$\tau_{2k}$ in Lemma \ref{dtauBC}, 
  \begin{equation}\label{momentmap}
  \mu(\tau_{2k})=\displaystyle{\frac{\displaystyle{\sum_{1\le i_1<\cdots<i_{2k}\le 2n} |{\rm pf}(i_1,\ldots,i_{2k})E_{i_1}\cdots E_{i_{2k}}|^2({\mathcal L}_{i_1}+\cdots+{\mathcal L}_{i_{2k}})}}{\displaystyle{\sum_{1\le i_1<\cdots<i_{2k}\le 2n}|{\rm pf}(i_1,\ldots,i_{2k})E_{i_1}\cdots E_{i_{2k}}|^2}}}\,.
     \end{equation}
Here ${\mathfrak h}_{\mathbb R}^*$ is the dual space of Cartan subalgebra of ${\mathfrak{sl}}(2n,{\mathbb R})$,
\[
{\mathfrak h}_{\mathbb R}^*={\rm Span}_{\mathbb R}\left\{{\mathcal
  L}_1,\ldots,{\mathcal L}_{2n}: \sum_{i=1}^{2n}{\mathcal
  L}_i=0\right\} \cong \mathbb{R}^{2n-1}\,,
\]
with the weights ${\mathcal L}_k$. The moment map $\mu$ can be extended to the map
over the product space of ${\mathbb P}(\wedge^{2k}{\mathbb R}^{2n})$ by
\begin{equation}\label{momentmaps}
\begin{array}{ccccc}
\mu&:&{\mathbb P}(\wedge^2{\mathbb R}^{2n})\times\ldots\times{\mathbb P}(\wedge^{2n-2}{\mathbb R}^{2n})&\longrightarrow & {\mathfrak h}_{\mathbb R}^*\\
  & & (\,\tau_2\,,\,\ldots\,,\,\tau_{2n-2}\,)&\longmapsto & \displaystyle{\sum_{k=1}^{n-1}\mu(\tau_{2k})}\,.
  \end{array}
\end{equation}
The moment polytope given by the image of the moment map is expressed by
the wedge product representation with the highest weight (see e.g. \cite{fulton:91}),
\[
w_*:=\sum_{k=1}^{n-1}\left( {\mathcal L}_1+\cdots+{\mathcal L}_{2k}\right)=\sum_{k=1}^{n-1} (n-k)({\mathcal L}_{2k-1}+{\mathcal L}_{2k})\,.
\]
This is a tensor product of the basic representations of $SL(2n)$, and it is customary denoted by $\Gamma_{0,1,0,\ldots,1,0}$ (see \cite{fulton:91}): In general, $\Gamma_{a_1,\ldots,a_{2n-1}}$ represents the irreducible representation of $SL(2n)$ with the highest weight $w_*=\sum_{k=1}^{2n-1} a_k({\mathcal L}_1+\cdots+{\mathcal L}_k)$.
The vertices of the polytope are given by the weights which are parametrized by the set of
numbers $\{0,1,\ldots,n-1\}$, i.e.
\[
w^{\alpha}=(\alpha_1,\alpha_2,\ldots,\alpha_{2n}):=\sum_{k=1}^{2n}\alpha_k{\mathcal L}_k\,.
\]
with $\alpha_k\in\{0,1,\ldots,n-1\}$, and each number of the set appears exactly twice in $\{\alpha_k\}$,
e.g. $w_*=(n-1,n-1,n-2,n-2,\ldots,1,1,0,0)$.
Each vertex of the polytope can be parametrized by a unique element of $W^P$, the set of
minimal representatives of the cosets in,
\[
W^P:=S_{2n}/W_P \quad {\rm with}\quad W_P=\langle s_1,s_3,\ldots,s_{2n-1}\rangle\,,
\]
where $s_k$ are the simple reflections, i.e. $s_k=(k,k+1)$ for $k=1,\ldots,2n-1$.
Note here that $W_P\cong \overbrace{S_2\times\cdots\times S_2}^n$ and
$|W^P|=(2n)!/2^n$.
For example, for $n=2$ and $W_P=\langle s_1,s_3\rangle$, we have
\[
W^P=\{\,e, \,s_2, \,s_1s_2, \,s_3s_2,\,s_1s_3s_2, \,s_2s_1s_3s_2\,\}\,.
\]
We also note that the quotient $W^P$ parametrizes the Bruhat decomposition of
$SL(2n)/P_{2\times 2}$, i.e. $SL(2n)/P_{2\times 2}=\sqcup_{w\in W^P}NwP_{2\times 2}$
where $N$ is a unipotent subgroup of lower triangular matrices.

  From (\ref{momentmap}), one can also see that those vertices are identified with the fixed points
  of the Pfaff lattice.  The identification is given as follows: For the vertex of $(\alpha_1,\ldots,\alpha_{2n})$, assign $\alpha_k$'s so that
  \[
  \alpha_{i_{2k+1}}=\alpha_{i_{2k+2}}=n-k-1\quad\quad k=0,1,\ldots,n-1\,.
  \]
Then the corresponding fixed point is given by the block upper triangular matrix $\hat{L}$ obtained in Theorem \ref{fixedL}.
 For example, the vertex $w^{\alpha}=(0,1,2,1,0,2)$ in the case of $n=3$ corresponds
to the matrix $\hat L$ given by
\[
\hat L=\begin{pmatrix}
0  &  1  &       &         &       &         \\
-z_3z_6 & z_3+z_6 &    1  &      &       &     \\
     &     &   -(z_3+z_6)  &    1   &     &       \\
     &     &  -(z_3+z_6+z_2)(z_3+z_6+z_4) & -(z_1+z_5) &   1  &     \\
     &    &              &         &     z_1+z_5    &     1     \\
     &    &               &       &  -  z_1z_5  &      0     
     \end{pmatrix}
 \]
where we have used ${\rm tr}(\hat L)=\sigma_1(1,\ldots,6)=0$, and all other entries are zero.
This fixed point can be parametrized by the element $s_1s_2s_1s_3s_5s_4s_3s_2\in S_{6}/\langle 
s_1,s_3,s_5\rangle$.

We summarize the result in this section as follows:
\begin{Theorem}\label{isospectralPfaff}
For the generic Pfaff lattice,
the image of the moment map (\ref{momentmaps}) is a convex polytope whose vertices
are parametrized by the set $W^P$ of minimal representatives of the
cosets in $S_{2n}/W_P$ with
$W_P=\langle s_1,s_3,\ldots,s_{2n-1}\rangle$. Those vertices are identified as the fixed points of the Pfaff flow given in Theorem \ref{fixedL}.
\end{Theorem}
 
\begin{Example}\label{n2momentP} 

\begin{figure}[t]
\includegraphics[width=10.5cm]{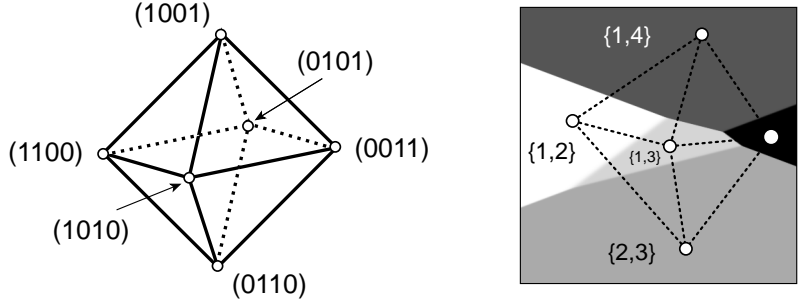}
\caption{The momentum polytope for the GOE-Pfaff lattice with $n=2$.
The right figure shows the function $b_1$ of the Pfaff lattice in $(t_1,t_2,t_3)$ with $t_3=-10$ (see Figure \ref{2sol:fig}).
Each asymptotic region of the $b_1$ is identified by
the vertex of the polytope.}
\label{n2poly:fig}
\end{figure}

In Example \ref{n2example}, we considered the Pfaff lattice for the GOE model in the case of
$n=2$. The momentum polytope associated to this model is given by the irreducible representation
with the highest weight $w_*={\mathcal L}_1+{\mathcal L}_2$, which is an octahedron. In Figure \ref{n2poly:fig}, we show the momentum polytope and explain how the polytope can be realized
from the solution of the Pfaff lattice in the space of $(t_1,t_2,t_3)$. One should note that the polytope
is just a dual diagram of the pattern of the solution $b_1$ in $(t_1,t_2,t_3)$, that is, identify
each region for constant $b_1$ as a vertex of the polytope, and the line between two values of $b_1$
as an edge of the polytope, and so on. The vertices are parametrized by the elements of $W^P$, i.e.
\[\begin{array}{ccc}
e=\{1,2\},\quad s_2=\{1,3\},\quad s_1s_2=\{2,3\},\quad s_3s_2=\{1,4\},\\
\quad s_1s_3s_2=\{2,4\}, \quad s_2s_1s_3s_2=\{3,4\}.
\end{array}
\]

The generic orbit can be described by a curve inside of the polytope approaching to
$\{1,2\}$ vertex as $t_1\to-\infty$ and to $\{3,4\}$ vertex as $t_1\to\infty$. Those vertices represent
the highest and lowest weights. The edges of the polytope correspond to non-generic orbits
of the Pfaff lattice. For example, the edge between  $\{1,2\}$ and $\{1,3\}$ is a solution
given by the $\tau_2$ function with the $B$-matrix, 
\[B=\begin{pmatrix}
0 & 1 & 1 & 0 \\
 & 0 & 0 & 0 \\
 &  &  0  &  0\\
 &   &     &   0
 \end{pmatrix}.
 \]
 Note that the $\tau_4=0$ for this choice of $B$-matrix, and the $\tau_2$ satisfies the KP equation.
The faces of the polytope  correspond to non-generic orbits (i.e. KP solutions).
Some explicit solutions are studied in \cite{kodama:06}.
We then note that a classification of the solutions of the Pfaff lattice can be obtained by
the representations of the $SL(2n)$ orbit in ${\mathbb P}(\Gamma_{0,1,0,\ldots,1,0})$.
\end{Example}

\subsection{The moment polytope for the GSE model \label{GSEpolytope}}
Recall that the $\tau$-functions for the GSE-Pfaff lattice are given by (\ref{tauGSE}), i.e.
\[
\tau_{2k}=\sum_{1\le i_1<\cdots<i_k\le n}\Delta(i_1,\ldots,i_k)^4E_{i_1}^2\cdots E_{i_k}^2\,,
\]
where $\Delta(i_1,\ldots,i_k)=\prod_{1\le j<l\le k}(z_{i_l}-z_{i_j})$. Writing those in the form,
\[
\tau_{2k}=\sum_{1\le i_1<\cdots<i_k\le n}\Delta(i_1,\ldots,i_k)^3F(i_1,\ldots,i_k)\,,
\]
with 
\[
F(i_1,\ldots,i_k):=2^{-\frac{k(k-1)}{2}}{\rm det}\begin{pmatrix}
F_{i_1}  &   F_{i_2}  &\cdots  &  F_{i_k}  \\
F'_{i_1} &  F'_{i_2} &  \cdots & F'_{i_k}  \\
\vdots  &  \vdots   &  \ddots  &  \vdots  \\ 
F^{(k-1)}_{i_1}&F^{(k-1)}_{i_2}&\cdots &F^{(k-1)}_{i_k}
\end{pmatrix}
\]
where $F_k=E^2_k=e^{2\theta(t,z_k)}$ and $F^{(j)}_k=\partial^j F_k/\partial t_1^j$. Since $F(i_1,\ldots,i_k)$ represents a coordinate
for $\wedge^k{\mathbb R}^n$, the $\tau_{2k}$ can be considered as
a point in ${\mathbb P}(\wedge^k{\mathbb R}^n)$ as discussed in the previous section.
Now we consider the moment map $\mu:{\mathbb P}(\wedge^k{\mathbb R}^n)\to {\mathfrak h}^*_{\mathbb R}$ where ${\mathfrak h}^*_{\mathbb R}$ is the dual space of Cartan subalgebra
of ${\mathfrak {sl}}(n,{\mathbb R})$,
\[
  \mu(\tau_{2k})=\displaystyle{\frac{\displaystyle{\sum_{1\le i_1<\cdots<i_{k}\le n} |{\Delta}(i_1,\ldots,i_{k})^3F_{i_1}\cdots F_{i_{k}}|^2({\mathcal L}_{i_1}+\cdots+{\mathcal L}_{i_{k}})}}{\displaystyle{\sum_{1\le i_1<\cdots<i_{k}\le n}|{\Delta}(i_1,\ldots,i_{k})^3F_{i_1}\cdots F_{i_{k}}|^2}}}\,,
  \]
where ${\mathcal L}_k$ are the weights of ${\mathfrak h}^*$ of ${\mathfrak{sl}}(n,{\mathbb R})$.
As in the previous case, the moment map $\mu$ can be extended to the product space,
\begin{equation}\label{GSEmoment}
\begin{array}{ccccc}
\mu&:&{\mathbb{RP}}^{n-1}\times{\mathbb P}(\wedge^2{\mathbb R}^{n})\times\ldots\times{\mathbb P}(\wedge^{n-1}{\mathbb R}^{n})&\longrightarrow & {\mathfrak h}_{\mathbb R}^*\\
  & & (\tau_2,\tau_4,\ldots,\tau_{2n-2})&\longmapsto & \displaystyle{\sum_{k=1}^{n-1}\mu(\tau_{2k})}\,.
  \end{array}
\end{equation}
Then the moment polytope is the permutahedron of $S_n$ (denoted as
$\Gamma_{1,\ldots,1}$ of the basic representation
of $SL(n)$ \cite{fulton:91}), which is the weight polytope with
the highest weight,
\[
w_*=\sum_{k=1}^{n-1}({\mathcal L}_1+\cdots+{\mathcal L}_k)=(n-1,n-2,\ldots,1,0)\,.
\]
Each vertex of the polytope is parametrized by the element of $S_n$. We summarize the result:
\begin{Theorem}\label{GSEvariety}
For the GSE-Pfaff lattice, the image of the moment map (\ref{GSEmoment}) is given by the permutahedron of the symmetric group $S_n$, whose vertices are the fixed points of the
flow.
\end{Theorem}

\begin{figure}[t]
\includegraphics[width=10.5cm]{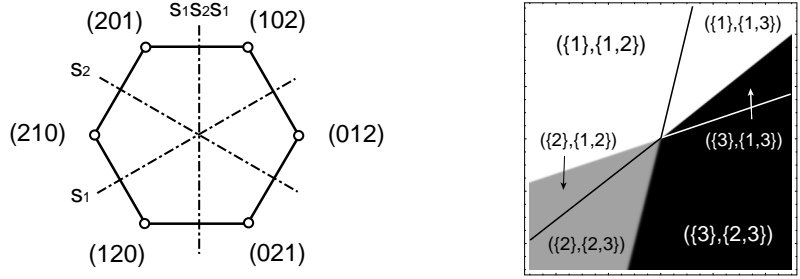}
\caption{The momentum polytope for the GSE-Pfaff lattice with $n=3$. The right figure shows the combined graph of $b_1$ and $b_2$ in Figure \ref{gse:fig}. Each $(\{i\},\{j,k\})$
indicates the set of asymptotic values $(b_1,b_2)$, i.e 
$b_1=z_i$ and $b_2=z_j+z_k$.
The left figure shows the moment polytope with the weights $(\alpha_1,\alpha_2,\alpha_3)=\alpha_1{\mathcal L}_1+\alpha_2{\mathcal L}_2+\alpha_3{\mathcal L}_3$, and $\alpha_k$ indicates the number of $k$ appearing in the set $(\{i\},\{j,l\})$. The Weyl elements $s_1, s_2$ and $s_1s_2s_1$
correspond to the boundaries of the regions $(\{i\},\{j,k\})$.}
\label{GSEpoly:fig}
\end{figure}

\begin{Example}  The momentum polytope for $n=3$ GSE model:
In Figure \ref{gse:fig}, the dominant exponential terms in the $\tau$-functions are plotted
in $t_1$-$t_2$ plane. The combined graph of $b_1=\partial\ln\tau_2/\partial t_1$ and $b_2=
\partial\ln\tau_4/\partial t_1$ divides the $t_1$-$t_2$ plane into 6 regions, each of which
is marked by the number set $(\{i\},\{j,k\})$ with $i=j$ or $i=k$ $(j<k)$ (see Figure \ref{GSEpoly:fig}).
The dual graph of the combined graph gives a hexagon as shown in
Figure \ref{GSEpoly:fig}, which is the moment polytope for the symmetric group $S_3$.
The edges of the hexagon correspond to non-generic flows, e.g. the edge between the vertices $(201)$
and $(102)$ corresponds to the flow given by the $\tau$-function with $c_2=0$,
\[
\tau_2=c_1E_1^2+c_3E_3^2\,.
\]
This then gives $\tau_4=c_1c_3(z_1-z_3)^4E_1^2E_3^2$, and the Pfaff lattice is reduced to
the model with $n=2$, i.e. a subsystem of the original lattice.

\end{Example}

 
\bibliographystyle{amsalpha}

\end{document}